\documentclass[letterpaper]{article}

\usepackage[english]{babel}
\usepackage[T1]{fontenc}

\usepackage[margin=1in]{geometry}

\usepackage[authoryear, comma]{natbib}

\usepackage{amsmath}
\usepackage{amssymb}
\usepackage{bm}

\usepackage{graphicx}
\usepackage{subcaption}
\usepackage{float}

\usepackage{indentfirst} 
\usepackage{tabu}

\usepackage{fancyhdr}
\usepackage{totpages}
\usepackage{hyperref}
\hypersetup{
    colorlinks=true,
    linkcolor=blue,
    citecolor=blue,
    filecolor=magenta,      
    urlcolor=cyan,
    breaklinks=true
}

\DeclareMathOperator{\sgn}{sgn}

\title{Computational Analysis of Interfacial Dynamics in Angled Hele-Shaw Cells: Instability Regimes}

\author{Daihui Lu, Federico Municchi\footnote{Present address: School of Mathematical Sciences, The University of Nottingham, University Park, Nottingham, NG7 2RD, UK.}~~and Ivan C.\ Christov\footnote{To whom the correspondence should be addressed. \href{mailto:christov@purdue.edu}{\texttt{christov@purdue.edu}}; \url{http://tmnt-lab.org}}\\[2mm]
\textit{School of Mechanical Engineering, Purdue University,}\\ \textit{West Lafayette, Indiana 47907, USA}}

\date{\today}

\begin{document}
\maketitle 

\begin{abstract}
We present a theoretical and numerical study on the (in)stability of the interface between two immiscible liquids, i.e., viscous fingering, in angled Hele-Shaw cells across a range of capillary numbers ($Ca$). We consider two types of angled Hele-Shaw cells: diverging cells with a positive depth gradient and converging cells with a negative depth gradient, and compare those against parallel cells without a depth gradient. A modified linear stability analysis is employed to derive an expression for the growth rate of perturbations on the interface and for the critical capillary number ($Ca_c$) for such tapered Hele-Shaw cells with small gap gradients. Based on this new expression for $Ca_c$, a three-regime theory is formulated to describe the interface (in)stability: (i) in Regime I, the growth rate is always negative, thus the interface is stable; (ii) in Regime II, the growth rate remains zero (parallel cells), changes from negative to positive (converging cells), or from positive to negative (diverging cells), thus the interface (in)stability possibly changes type at some location in the cell; (iii) in Regime III, the growth rate is always positive, thus the interface is unstable. We conduct three-dimensional direct numerical simulations of the full Navier--Stokes equations, using a phase field method to enforce surface tension at the interface, to verify the theory and explore the effect of depth gradient on the interface (in)stability. We demonstrate that the depth gradient has only a slight influence in Regime I, and its effect is most pronounced in Regime III. Finally, we provide a critical discussion of the stability diagram derived from theoretical considerations versus the one obtained from direct numerical simulations. 

\end{abstract}


\section{Introduction}


Secondary and enhanced oil recovery (EOR) require the stable displacement of flowing subsurface hydrocarbons by an injected fluid \citep{L89}. Water, being abundant, is a typical choice of displacing fluid. However, it is well known that this displacement process is unstable due to the lower viscosity of water (even with various surfactants and additives mixed into it) compared to immiscible subsurface hydrocarbons (e.g., crude oil) \citep{Hill1952,Saffman1958,Chuoke1959}. Consequently, ``the average oil recovery factor worldwide is only between 20\% and 40\%'' even accounting for modern EOR techniques \citep{Metal14}. Therefore, there is an ongoing need to better understand fluid--fluid interface instabilities in the presence of a viscosity contrast and surface tension. An additional complication, beyond the physicochemical properties of the fluids used in EOR, is that subsurface formations, from which hydrocarbons are to be extracted, are naturally heterogeneous with geometric variations in the flow passages \citep{JMY2003}. In this regard, recent experiments \citep{Al-Housseiny2012}, which employed the Hele-Shaw analogy for modeling the conductivity of shaped fractures \citep{ZB96}, have re-emphasized the importance of understanding the effect of geometric on the stability of fluid--fluid interfaces in the subsurface. 

A similar problem arises during modern hydraulic fracturing (``fracking'') processes as well. Although the effectiveness of fracking was demonstrated in 1947 \citep{EN00}, it remains a challenging approach to energy production, in particular, due to the complex thermo-hydro-mechanical-chemical coupled processes involved across multiple space and time scales \citep{Hyman2016}. In this case, the displacing fluid, primarily water with proppants \citep{YW2014,O17}, is injected into a wellbore at high pressure to create cracks in the subsurface rock formation. The natural heterogeneity and geometric variations of flow passages in these formations thus become even more pronounced during fracking. Once again, the instability of the interface between the fracking (displacing) fluid and the hydrocarbons (defending, or displaced, fluid) must be managed to ensure a high oil recovery rate, especially during overflushing \citep{OBZD18}. As with EOR, it is desirable to displace a stable interface through the fractured rock so as to produce a ``clean'' sweep of the fracture, minimizing the oil or gas film layers left behind. Thus, there is an impetus to study fluid--fluid interface instabilities in complex/variable geometries, which is subject of the present work.

\subsection{Context for studying interfacial instabilities}

Interfacial instabilities characterized by the competition of surface tension and flow are quite common in nature and industry: the formation of snow flakes during solidification of a liquid \citep{Langer1980}, ``ribbing'' in coating flows due to the variable gap between two rollers \citep{Weinstein2004}, breakup of liquid threads confined in microfluidic devices for the generation of emulsions \citep{SLA2016} or for inkjet printing \citep{BGB13}, and even in electromechanical systems such as flow batteries \citep{CHH16}, to list a few. In most (but not) all of these examples, the instabilities are caused by a viscosity contrast at the fluid--fluid interface \citep{Homsy1987}. Finger-like patterns form as the unstable interface grows, which has led to this phenomenon being termed \emph{viscous fingering} \citep{Saffman1986}. 

In the 1950s, \citet{Hill1952}, \citet{Saffman1958}, and \citet{Chuoke1959} laid the foundations for the study of viscous fingering through both theoretical analysis and experiments. Specifically, \citet{Hill1952} performed a one-dimensional (1D) stability analysis and conducted quantitative experiments for both stable and unstable interfaces between sugar liquors and water.  \citet{Saffman1958} considered a less viscous fluid (air) displacing a more viscous one (glycerine) in a Hele-Shaw cell, i.e., a thin gap between two closely spaced flat plates, and predicted the finger's growth rate via \emph{linear stability analysis}. \citet{Saffman1958} additionally predicted and verified that when a single finger forms in a Hele-Shaw cell, it occupies almost exactly half the width of the cell, for most experiments, which was later supported by the theory of \citet{pitts80}. Since then, Saffman and Taylor's approach has enabled a significant amount of theoretical and experimental research on interfacial instabilities. 

\subsection{Control of interfacial instabilities}
\citet{P60} (as well as a parallel work by \citet{PG61}) analyzed the effect of a gap gradient on the Saffman--Taylor problem \citep{Saffman1958}, in the context of thin films between rollers and spreaders. More recently, however, this problem has received renewed attention \citep{Al-Housseiny2012} due to the analogy between flow in a Hele-Shaw cell and flow through shaped subsurface fractures \citep{ZB96} as well as flows in certain porous media \citep{Bear1972,Homsy1987}.

Specifically, due to the importance of interfacial instabilities (such as viscous fingering) in confined geometries across many applications, attempts to control them via stabilizing gradients have been made \citep{M02}. To this end, the linear stability analysis in the presence of a gap gradient (i.e., in an angle Hele-Shaw cell) was re-interpreted as a possible stabilization (control) mechanism \citep{Al-Housseiny2012}. \citet{Zhao1992} previously revisited Pearson's problem experimentally and \citet{Dias2010} had updated the linear stability analysis to predict tip-splitting in the presence of variations of the flow passage. 

Since then, an explosion of works has analyzed a variety of related interfacial instability problems in ``non-standard'' Hele-Shaw configurations \citep{MMM19}, both fixed geometries \citep{DJ2013,Al-Housseiny2012,Al-Housseiny2013,hws16,Jackson2017,GSW17,BT18} and those with flow-driven geometric changes \citep{PPIHJ12,ahcs13,PPPRJH13}. Further variants on the same problem also include controlling the injection flow rate \citep{DM2010,DPM2010,DALC2012}, changing the permeability by adjusting the structure of the porous medium \citep{Jackson2017,Rabbani2018,Brandao2018}, applying an external force via rotation of the geometry or through a magnetic field \citep{CMC1996,ALPC2003,MAL2005,ALM2018}, changing the fluid properties through the viscosity ratio \citep{ADM2017}, using non-Newtonian fluids \citep{VM2000,LBP2002,PT2007,BOD15} or even adding a suspended particulate phase \citep{XKL2016,KXL2017}.

We are interested in geometric controls. To this end, there are three primary ways to alter the physical geometry of an experimental Hele-Shaw apparatus: (i) creating a gradient along the flow direction by relaxing the requirement that the plates be parallel \citep{Zhao1992,Dias2010,Al-Housseiny2012,Al-Housseiny2013,BT18,ADM18,MMM19}; (ii) using an elastic membrane (that deforms due to flow underneath it) instead of a solid top plate \citep{PPIHJ12,ahcs13,PPPRJH13}; and (iii) lifting one of the plates in a time dependent manner \citep{DM2010,zks2015,DPPCCO17}. Among these possibilities, the case of a geometric gradient in the flow direction has attracted special attention because it naturally imitates the non-uniform, fractured subsurface flow passages \citep{Metal14,O17}. The gradient could be (a) positive for an increasing gap depth in the flow direction (termed a \emph{diverging} Hele-Shaw cell), or (b) negative for a decreasing gap depth (termed a \emph{converging} cell).

\subsection{Goals and outline of this work}
Currently, despite extensive research on the topic, the predictions of mathematical analysis of the interfacial stability in tapered geometries (or attendant 2D numerical simulations) have not been verified through three-dimensional (3D) direct numerical simulation (DNS) of flow and interfacial instability. In fact, a study by \citet{DYL11} concluded that a ``3D [numerical] model is preferred to obtain a better comparison with experimental results.'' Indeed, in 3D DNS, unlike physical experiments or the numerous previous simulations of the depth-averaged Hele-Shaw equations, we have control over the entire problem setup, which allows us to capture the full physics of the problem. The goal of this paper is to fill this knowledge gap for rectangular Hele-Shaw cells with nonuniform gap thickness by: (i) extending the linear stability theory of interfacial instability in Hele-Shaw cell with clear fluids by taking into account the local streamwise variation of parameters (e.g., capillary number, depth of the cell, and so on); (ii) supplementing and verifying the theoretical analysis with ``full'' 3D DNS.

To this end, this paper is organized as follows. In Sec.~\ref{sec:linear}, we derive the (growth) decay rate of a (un)stable fluid--fluid interface between two immiscible phases in a Hele-Shaw with variable gap thickness (but constant gap gradient) via linear stability theory. Specifically, starting with Darcy's equation (i.e., the depth-averaged momentum equation) and the continuity equation, we obtain a Laplace equation for pressure. By solving the latter, we find the pressure jump at the fluid--fluid interface, and we match this pressure jump to the one found from the Young--Laplace equation. Thus, we arrive at the growth rate of the interface. On the basis of this mathematical result, we then classify the interface stability into three flow regimes depending on the difference between a critical capillary number and inlet or outlet capillary numbers, generalizing previous work on this problem by  \citet{Al-Housseiny2012}. Next, in Sec.~\ref{sec:verification}, we perform a series of DNSs (the methodology for which is described in Sec.~\ref{sec:simulations}) in a specific set of Hele-Shaw geometries. We use the simulations to verify our mathematical model (i.e., the theory developed in Sec.~\ref{sec:linear}) for the growth rate of the interface. Then, in Sec.~\ref{sec:gradient}, we discuss the effect of the gap gradient on the stability based on the theoretical solution and further numerical experiments. In Sec.~\ref{sec:sd}, we compare and discuss the flow regimes maps (in the 2D parameter space defined by the capillary number and the gap gradient) determined by theoretical and numerical analyses. Finally, conclusions stated and avenues for future work are discussed in Sec.~\ref{sec:conclusion}.

\section{Linear stability analysis}
\label{sec:linear}

\begin{figure}[ht]
\centering
\includegraphics[width=0.8\textwidth]{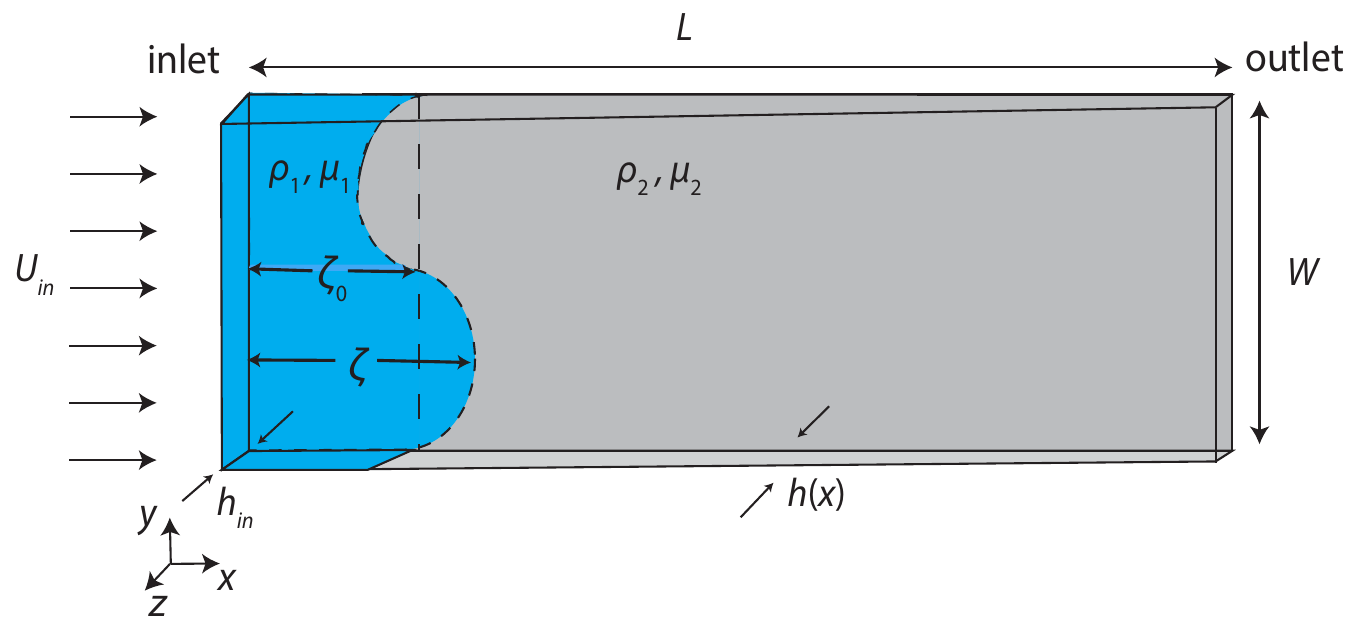}
\caption{Top view schematic configuration of the flow in a rectilinear Hele-Shaw cell with a constant depth gradient $\alpha$. The depth of the cell, $h(x)=h_{in}+\alpha x$, satisfies $\max_x h(x) \ll W $ and $\max_x h(x) \ll L$ and $h_{in}$ is the depth at the inlet.  Although we neglect gravity, in this view it would act in the negative $z$-direction. The flow is in the $x$-direction, and the Hele-Shaw cell's gap thickness, $h(x)$, only varies in this direction.}
\label{fig:schematic_figure}
\end{figure}

Consider two immiscible and incompressible viscous fluids flowing in a narrow gap between two rigid plates (see Fig.~\ref{fig:schematic_figure}) with a constant depth gradient $\alpha$. The depth of the cell $h(x)=h_{in}+\alpha x$ satisfies $\max_x h(x) \ll W $ and $\max_x h(x) \ll L$.  Although we neglect gravity, in Fig.~\ref{fig:schematic_figure} it would act in the negative $z$-direction. The flow is in the $x$-direction, and the Hele-Shaw cell's gap thickness, $h(x)$, only varies in this direction. The densities and viscosities of the displacing and defending fluids are denoted respectively as \(\rho_1\), \(\mu_1\) and \(\rho_2\), \(\mu_2\).  A fully developed flow of fluid 1 (the displacing fluid), with an average (in the $y$-$z$ cross-section) velocity $U_{in}$, pushes into a quiescent fluid 2 (the defending fluid). Between the two fluids there exists an interface that, due to immiscibility, is endowed with surface tension \(\gamma\). The interface is not necessarily flat, and its shape is given by \(x = \zeta(y,t)\). The horizontal direction perpendicular to the flow, i.e., the $y$-direction in Fig.~\ref{fig:schematic_figure}, is assumed to be large compared to the typical gap size. Therefore, consistent with the linear stability analysis to be carried out below (and also previous work of \citet{Al-Housseiny2012,Miranda1998}), we consider the interface to be periodic in $y$. Specifically, we shall apply a full-period  initial perturbation to an initially flat interface to respect the periodic boundary conditions (PBCs) at $y = 0$ and $y = W$. Experiments have shown that PBCs have a similar effect to physical sidewalls in a cylindrical Hele-Shaw cell \citep{ZM1990}, i.e., two coaxial cylinders separated by a small gap.

\subsection{Linear growth rate}

Following \citet{Al-Housseiny2013}, we start our linear stability analysis with the 2D (i.e., depth- or $z$-averaged) governing equations for the viscous fluid flow in the thin gap between two closely spaced plates \citep{Panton}:
\begin{equation}
\bm{u}_j = -\frac{h^2}{12 \mu_j} \nabla p_j,
\label{eq:Darcy}
\end{equation}
where the subscript $j=1,2$ represents the displacing and defending fluid, respectively; $\bm{u}_j$ is the depth-averaged velocity field of fluid $j$ in the $(x,y)$ plane, $p_j$ is the pressure field of fluid $j$, while $h$ and $\mu_j$ were defined above (see also Fig.~\ref{fig:schematic_figure}). Equation~\eqref{eq:Darcy} is supplemented by the continuity (conservation of mass) equation for each incompressible fluid phase:
\begin{equation}
\nabla \cdot(h\bm{u}_j) = 0.
\label{eq:cont}
\end{equation}
The flow has been assumed to be fully developed and steady.

Substituting Eq.~\eqref{eq:Darcy} into Eq.~\eqref{eq:cont}, we obtain the governing equation for the pressure in each fluid
\begin{equation}
\nabla ^2 p_j + \frac{3\alpha}{h(x)}\frac{\partial p_j}{\partial x} = 0.
\label{eq:govern}
\end{equation}
In an angled Hele-Shaw cell, the depth is $h(x) = h_{in} + \alpha x$, so
\begin{equation}
\frac{1}{h(x)} = \frac{1}{h_{in}+\alpha x}= \frac{1}{h_{in}}-{\frac {\alpha x}{h_{in}^{2}}}  + \mathcal{O}( \alpha^2) . \label{eq:h}
\end{equation}
Substituting Eq.~\eqref{eq:h} back into Eq.~\eqref{eq:govern}, and neglecting the $\mathcal{O}(\alpha^2)$ terms (consistent with the lubrication approximation under which Eq.~\eqref{eq:Darcy} is derived \citep[see, e.g.,][]{Panton}), the pressure equation now has constant coefficients:
\begin{equation}
\frac{\partial^2 p_j}{\partial x^2}+\frac{\partial^2 p_j}{\partial y^2}+\frac{3\alpha}{h_{in}}\frac{\partial p_j}{\partial x} = 0\qquad (\alpha \ll 1).
\label{eq:govern3}
\end{equation}

Next, assume a flat base state for the (unperturbed) interface shape, denoted $\zeta(y,t)=\zeta_0(t)$. Then, we express the perturbed interface as a Fourier series:
\begin{equation}
\zeta(y,t) = \zeta_0(t) + \sum_{n\ne 0} \epsilon_n a_n e^{ik_ny+\lambda_n (t)},
\label{eq:iface-perturb}
\end{equation}
which satisfies the condition of periodic boundary conditions (PBCs) in $y$. In Eq.~\eqref{eq:iface-perturb}, the time derivative of $\lambda_n(t)$, denoted henceforth as $\dot{\lambda}_n$ and not necessarily constant, represents the growth rate of mode $n$, and $k_n=2\pi n/W$ is its spatial wave number. We take $k_n>0$ without loss of generality. The magnitude of each perturbation mode is quantified by a dimensionless number $\epsilon_n$, and $\epsilon_{n'} \ne \epsilon_n$ for any two modes $n'$ and $n$ in a mode-coupling analysis \citep{Miranda1998}. In the present work, we restrict ourselves to a single mode analysis, thus the sum may be dropped:
\begin{equation}
\zeta(y,t) = \zeta_0(t) + \epsilon a e^{iky+\lambda (t)} .
\label{eq:iface-perturb2}
\end{equation}
This approximation is justified since all mode coupling terms would be at least second order, or more specifically, coupling of modes $n'$ and $n$ would be of order $\epsilon_{n'} \epsilon_n$, which would be dropped eventually in our linear stability analysis below.

Next, we expand each phase's pressure $p_j$ in perturbation series. Consistent with the interfacial perturbation in Eq.~\eqref{eq:iface-perturb2}, only terms up to $\mathcal{O}(\epsilon)$ are kept:
\begin{equation}
p_j(x,y,t) = p_{0j}(x;t) + \epsilon p_{1j}(x,y;t), 
\label{eq:pj}
\end{equation}
where $p_{0j}(x;t)$ represents the base state, i.e., the pressure drop across the channel under uniform displacement. Meanwhile, $p_{1j}(x,y;t)$ is the pressure perturbation due to the interfacial disturbance. Note that the perturbative pressure expansion from Eq.~\eqref{eq:pj} satisfies the steady PDE, i.e., Eq.~\eqref{eq:govern3}, but may depend on time as a parameter due to the fluid--fluid interface's motion. Consistent with these definitions, the pressure gradient of the base state $p_{0j}(x;t)$ satisfies Darcy's equation and $p_{1j}(x,y;t)$ must vanish away from the interface, i.e., $\lim\limits_{x \to - \infty} p_{11}(x,y;t) = \lim\limits_{x \to + \infty} p_{12}(x,y;t)= 0$.

We proceed by expressing the pressure perturbation as a Fourier series:
\begin{equation}
p_{1j}(x,y;t) = \sum_n g_{jn}(x) e^{ik_ny+\lambda_n (t)}, \quad \quad j=1,2,
\label{eq:p1}
\end{equation}
where each $g_{jn}=\mathcal{O}(1)$. Again, Eq.~\eqref{eq:p1} can be reduced to a single-mode representation due to higher-order terms being dropped in our linear stability analysis:
\begin{equation}
p_{1j}(x,y;t) =  g_{j}(x) e^{iky+\lambda (t)}.
\label{eq:p1_s}
\end{equation}

Solving the pressure equation by substituting Eqs.~\eqref{eq:pj} and \eqref{eq:p1_s} into Eq.~\eqref{eq:govern3} (see Appendix for details), we obtain the pressure jump across the fluid--fluid interface:
\begin{multline}
\big(p_1 - p_2\big)\big|_{x=\zeta(y,t)} = \frac{4 U\big(\zeta_0(t)\big) h_{in}}{ \alpha [h(\zeta_0(t))]^2}(\mu_1 - \mu_2) \\ + \epsilon e^{iky+\lambda (t)} \frac{a \gamma Ca}{[h\big(\zeta_0(t)\big)]^2} \left\{ (1-M) - \left[ \frac{\dot{\lambda}}{U\big(\zeta_0(t)\big)}  + \alpha \left(\frac{3}{h_{in}} - \frac{2}{h\big(\zeta_0(t)\big)} \right)  \right] \frac{1+M}{k} \right\} + \mathcal{O}(\epsilon^2)\qquad (\alpha \ll 1),
\label{eq:p2-p1}
\end{multline}
where $Ca := 12U \mu_2/\gamma$ is the definition of the capillary number, $\gamma$ is the interfacial surface tension as before, $M := \mu_1/\mu_2$ is defined as the ratio of the fluids' viscosities, and $U=U\big(\zeta_0(t)\big)$ and $h=h\big(\zeta_0(t)\big)$ are the local (non-constant) velocity and depth at the unperturbed interface, respectively. It follows that $Ca$ also depends on $t$, implicitly, through $U\big(\zeta_0(t)\big)$ and $h\big(\zeta_0(t)\big)$; however, we have left this dependency implicit to simplify the notation.

On the other hand, the capillary pressure jump at the interface $x=\zeta(y,t)$ also satisfies the Young--Laplace equation. If the defending fluid wets the wall, i.e., the contact angle between the defending fluid and the wall is $\theta_c=0$, then Park and Homsy's analysis \citep{PH84} yields
\begin{equation}
\big(p_1 - p_2\big)\big|_{x=\zeta(y,t)}  = \frac{2\gamma}{h[\zeta(y,t)]} \left(1+3.8 Ca^{2/3}+\cdots\right) + \frac{\gamma}{R} \left[\frac{\pi}{4} + \mathcal{O}(Ca^{2/3}) \right], 
\label{eq:p_jump_Homsy}
\end{equation}
where $R$ is the radius of curvature of the interface, and we have taken into account the variable gap depth, $h[\zeta(y,t)]\ne const.$, as suggested by \citet{MM95}. Neglecting higher-order terms (in this linear, $Ca \ll 1$ analysis), Eq.~\eqref{eq:p_jump_Homsy} simplifies to
\begin{equation} 
\big(p_1 - p_2\big)\big|_{x=\zeta(y,t)}  = \frac{2\gamma}{h[\zeta(y,t)]} + \frac{\pi}{4} \frac{\gamma}{R}. \label{eq:p_jump}
\end{equation}
If the wetting is not perfect, considering the contact angle $\theta_c$, the pressure jump is written as 
\begin{equation}
\big(p_1-p_2\big)\big|_{z=\zeta(y,t)} = \gamma\left\{\frac{2 \cos \theta_c}{h[\zeta(y,t)]}+ f(\theta_c) \kappa\right\},
\label{eq:YoungLap}
\end{equation}
where $\kappa=1/R$ is the curvature of the interface, defined as
\begin{equation}
\begin{split}
\kappa &:= - \frac{\partial^2 \zeta/\partial y^2}{[1+(\partial \zeta/\partial y)^2]^{3/2}} \\
& =  - \frac{\partial^2 \zeta}{ \partial y^2} \left[ 1 - \frac{3}{2} \left( \frac{\partial \zeta}{\partial y} \right)^2  + \mathcal{O} \left(  \left( \frac{\partial \zeta}{\partial y} \right)^4 \right)\right] \\
& = k^2 \epsilon a e^{iky+\lambda (t)} + \mathcal{O}(\epsilon^2),
\end{split}
\label{eq:curvature_expand}
\end{equation}
and $f(\theta_c)$ is a function of the contact angle to account for the interface curvature within the gap \citep{Homsy1987,PH84}. Specifically, based on the analysis of \cite{LGBK07}, 
\begin{equation}
f(\theta_c) = \left(\frac{\pi}{4} - \frac{\theta_c}{2}\right)\frac{1+\sin\theta_c}{\cos\theta_c}.
\end{equation}
Of course, this general expression yields the two standard cases: $f(0)=\pi/4$ and  $f(\pi/2) = 1$ (interpreted as a limit).

Now, using the expansion in Eqs.~\eqref{eq:h} and~\eqref{eq:curvature_expand}, Eq.~\eqref{eq:YoungLap} becomes
\begin{equation}
\begin{split}
\big(p_1 - p_2\big)\big|_{x=\zeta(y,t)}  = \frac{2\gamma \cos \theta_c}{h\big(\zeta_0(t)\big)} + \epsilon a \gamma e^{iky+\lambda (t)} \left\{ f(\theta_c)k^2 - \frac {2 \cos \theta_c \alpha }{[h\big(\zeta_0(t)\big)]^2}  \right\} + \mathcal{O}(\epsilon^2).
\label{eq:p2-p1_2}
\end{split}
\end{equation}
Matching the $\mathcal{O}(\epsilon)$ terms in Eqs.~\eqref{eq:p2-p1} and  \eqref{eq:p2-p1_2}, we obtain
\begin{multline}
e^{iky+\lambda (t)} \frac{Ca \gamma a}{[h\big(\zeta_0(t)\big)]^2} \left\{ (1-M) - \left[ \frac{\dot{\lambda}}{U\big(\zeta_0(t)\big)}  + \alpha \left(\frac{3}{h_{in}} - \frac{2}{h\big(\zeta_0(t)\big)} \right) \right] \frac{1+M}{k} \right\} = \\
\left\{ f(\theta_c) k^2-\frac{2\alpha \cos \theta_c}{[h\big(\zeta_0(t)\big)]^2}\right\} \gamma a e^{iky+\lambda (t)}.
\end{multline}
Rearranging the last equation, yields the \emph{final form} of our theoretical prediction for the growth rate $\dot{\lambda}(t)$ of an interface between immiscible fluids in a rectilinear Hele-Shaw cell with a constant depth gradient $\alpha$:
\begin{equation}
(1+M) \left[ \frac{\dot{\lambda}}{U\big(\zeta_0(t)\big)}  + \alpha \left(\frac{3}{h_{in}} - \frac{2}{h\big(\zeta_0(t)\big)} \right) \right] = \left(1-M+\frac{2\alpha \cos \theta_c}{Ca}\right) k-\frac{f(\theta_c) k^3 [h\big(\zeta_0(t)\big)]^2}{Ca} \qquad (\alpha \ll 1).
\label{eq:GR}
\end{equation}

Equation~\eqref{eq:GR} differs from the solution discussed by \citet{Al-Housseiny2013} in that it does not assume an instantaneous development of the instability; rather the dynamics occurs over a finite time span during which the interface ``sees'' the cell's depth variation. Thus, Eq.~\eqref{eq:GR} captures the dynamic interplay between the growth/decay of a perturbation and the flow-wise geometric variations it encounters, as exemplified by $h \big(\zeta_0(t)\big)$. Specifically, we have defined a non-constant $Ca = 12 U \mu_2/\gamma$, where $U=U\big(\zeta_0(t)\big)$ is \emph{not} the fixed inlet value but, rather, it is the average velocity at some downstream cross-section $x=\zeta_0(t)$. $U=U\big(\zeta_0(t)\big)$ can be easily determined by conservation of mass; see, e.g., the discussion of  Eq.~\eqref{eq:Ca} below. Consequently, $\lambda(t)$ is governed by an \emph{ordinary differential equation} (not an algebraic equation), and it may grow (or decay, or both) during its time evolution, despite the sign of $\dot{\lambda}$ obtained from a ``frozen-time'' linear stability analysis \citep{KK14}.

In the absence of a depth gradient, i.e., $\alpha = 0$ (thus, $U$ and $Ca=const.$), Eq.~\eqref{eq:GR} reduces to
\begin{equation}
\dot{\lambda}  = \left(\frac{1-M}{1+M}\right) U k -\frac{f(\theta_c) h_{in}^2 U}{ Ca (1+M)} k^3, \label{eq:GR3}
\end{equation}
which agrees exactly with the growth rate for the fingering instability given by Homsy~\citep{Homsy1987} for $\theta_c = \pi/2$, and if gravity and the Rayleigh--Darcy convection terms are neglected therein.

\subsection{Classification of instability regimes}

The threshold of instability is determined by setting $\dot{\lambda}=0$. Then, from Eq.~\eqref{eq:GR}, we obtain
\begin{equation}
3 \alpha (1+M)   = \left(1-M+\frac{2\alpha \cos \theta_c}{Ca}\right) h_{in}k-\frac{f(\theta_c) [h\big(\zeta_0(t)\big)]^2 h_{in}}{Ca} k^3 \qquad (\alpha \ll 1).
\label{eq:lambda0}
\end{equation}
Solving for the \emph{critical} capillary number, $Ca_c$, for fixed $k$ in Eq.~\eqref{eq:lambda0}, we obtain
\begin{equation}
Ca_c = \frac{2\alpha \cos\theta_c - f(\theta_c) k^2[h\big(\zeta_0(t)\big)]^2}{ (1+M) \left(\frac{3}{h_{in}} - \frac{2}{h\big(\zeta_0(t)\big)} \right) \frac{\alpha}{k} + (M-1)}.
\label{eq:Cac}
\end{equation}
This critical capillary number determines the threshold of instability for the fluid--fluid interface. The interface is stable if $Ca < Ca_c$; it is unstable if $Ca > Ca_c$. To the best of our knowledge, Eq.~\eqref{eq:Cac} is a new result because $Ca_c$ implicitly depends on the interface location $x=\zeta_0(t)$ through $h\big(\zeta_0(t)\big)$. Now, let us consider the possible flow and instability regimes in Hele-Shaw cells with a gap gradient by analyzing Eq.~\eqref{eq:Cac} in detail. Importantly, these flow regimes could not be determined (i.e., they do not exists) in previously analyses based on a fixed $Ca_c$.

\paragraph{Parallel cell.} In this case, there is no depth gradient, i.e., $\alpha = 0$, and Eq.~\eqref{eq:Cac} becomes
\begin{equation}
Ca_c = -\frac{f(\theta_c) k^2h_{in}^2}{(M-1)}.
\end{equation}
The capillary number is constant along the cell, i.e., $Ca(x) \equiv Ca_{in} = Ca_{out}$. So, the critical capillary number delineates three regimes when compared with the inlet capillary number:
\begin{itemize}
\item \textbf{Regime I}: $Ca_{in} < Ca_c$. In this regime, the growth rate is negative, $\dot{\lambda}<0$. The finger's growth is suppressed, and the interface becomes flat asymptotically.

\item \textbf{Regime II}: $Ca_{in} = Ca_c$. In this regime, the growth rate is zero, $\dot{\lambda}=0$. The finger's growth is neither inhibited nor triggered (i.e., this represents a marginally stable case). The finger's length remains constant, and linear stability cannot determine whether an initial perturbation will grow or decay.

\item \textbf{Regime III}: $Ca_{in} > Ca_c$. In this regime, the growth rate is positive, $\dot{\lambda}>0$. The  interface is unstable and the finger's growth is triggered.
\end{itemize}

\paragraph{Diverging cell.} In this case, $\alpha > 0$. The depth-averaged velocity of a stable flat interface is intrinsically a function of the channel's depth along the flow direction due to conservation of mass. This leads us to specifically decompose the capillary number into an inlet (constant) and local (variable) contribution:
\begin{equation}
    Ca = \frac{12\mu_2U}{\gamma} = Ca_{in}\frac{h_{in}}{h(\zeta_0)}, 
\label{eq:Ca}
\end{equation}
where $Ca_{in}=12\mu_2 U_{in}/\gamma$ and $h_{in}$ are the constant capillary number and depth at the inlet of the cell. $Ca$ as defined in Eq.~\eqref{eq:Ca} is implicitly a function of $t$ through $\zeta_0$. This local $Ca$ decreases with $x$ because the velocity is decreasing for a diverging cell (expanding cross-sectional area). Although $Ca_c$ decreases as well, according to Eq.~\eqref{eq:Cac}, as $h(\zeta_0)$ increases, the change of the local $Ca$ is faster than $Ca_c$. Therefore, we can still introduce three regimes by comparing the local $Ca$ with $Ca_c$: 
\begin{itemize}
\item \textbf{Regime I}: \(Ca_{in} < Ca_c\). In this regime, the growth rate is always negative, $\dot{\lambda}<0$, and the finger's growth is inhibited.

\item \textbf{Regime II}: \(Ca_{in} > Ca_c\) and \(Ca_{out} < Ca_c\). In this regime, the growth rate is positive, $\dot{\lambda}>0$, at the inlet, \emph{but} changes sign becoming negative, $\dot{\lambda}<0$, at some point downstream in the cell. Thus, the finger's length initially increases but then saturates and would be expected to decrease at longer times. 

\item \textbf{Regime III}: \(Ca_{out} > Ca_c\). In this regime, the growth rate is always positive, $\dot{\lambda}>0$, and the initial finger perturbation grows in time.
\end{itemize}

\paragraph{Converging cell.} In this case, $\alpha < 0$. Now, local capillary number from Eq.~\eqref{eq:Ca} increases along the flow direction. Again, three instability regimes can be defined by comparing the local $Ca$ with $Ca_c$: 
\begin{itemize}
\item \textbf{Regime I}: \(Ca_{out} < Ca_c\). In this regime, the growth rate is always negative, $\dot{\lambda}<0$, and the finger's growth is inhibited.

\item \textbf{Regime II}: \(Ca_{in} < Ca_c < Ca_{out} \).  In this regime, the growth rate is negative, $\dot{\lambda}<0$, at the inlet, \emph{but} changes sign becoming positive, $\dot{\lambda}>0$, at some point downstream in the cell. The finger's length decreases initially but grows at longer times. 

\item \textbf{Regime III}: \(Ca_{in} > Ca_c\). In this regime, the growth rate is always positive, $\dot{\lambda}>0$, and the initial finger perturbation continues to grow in time.
\end{itemize}

It is important to note that this analysis predicts that, in diverging and converging cells, Regime II exists for a \emph{finite} range of $Ca_{in}$ values. This observation is in stark contrast to the parallel cells for which Regime II is simply the marginally stable case $Ca_{in}=Ca_c$. Thus, in diverging and converging Hele-Shaw cells, the stability of a perturbation may change during its evolution. Previous studies of elastic-walled Hele-Shaw cells also commented on this effect \citep[\S4.2.1]{PPPRJH13}.

\begin{figure} [ht]
  \centering
    \begin{subfigure}{0.48\textwidth}
    \includegraphics[width=\textwidth]{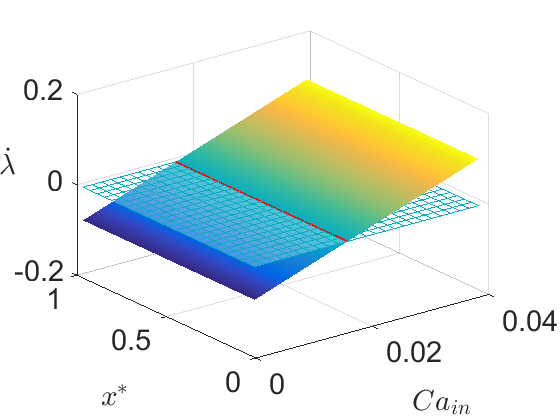}
    \subcaption{a parallel cell ($\alpha = 0$)}
    \label{fig:GR_para}
  \end{subfigure}
  \hfill
  \begin{subfigure}{0.48\textwidth}
    \includegraphics[width=\textwidth]{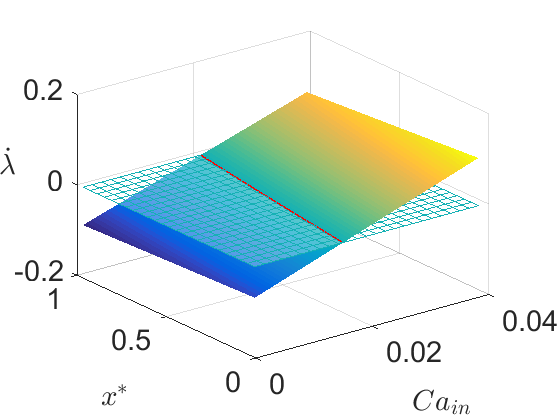}
    \subcaption{a diverging cell ($\alpha = 5 \times 10^{-4}$)}
    \label{fig:GR_div}
  \end{subfigure}
\begin{subfigure}{0.48\textwidth} 
    \includegraphics[width=\textwidth]{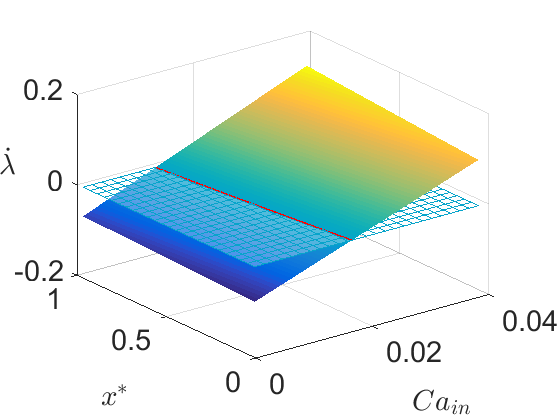}
    \subcaption{a converging cell ($\alpha = -5 \times 10^{-4}$)}
    \label{fig:GR_conv}
  \end{subfigure}
\caption{The growth rate $\dot\lambda$ as a function of inlet capillary number $Ca_{in}$ and the flow-wise dimensionless position $x^* = x/L$ in different cells: (a) a parallel cell, (b) a diverging cell, and (c) a converging cell, with $M = 0.0154$ and other geometric/material properties as in Table~\ref{tab:property}. The intersection of the growth rate surface (shaded) and the $\dot{\lambda}=0$ surface (meshed) is marked by red lines, indicating the change of sign of $\dot{\lambda}$ and an exchange of stability.}
\label{fig:GR}
\end{figure}

Now, to illustrate these three regimes for parallel, diverging and converging Hele-Shaw cells, in Fig.~\ref{fig:GR}, we plot the growth rate $\dot{\lambda}$ as a function of the inlet capillary number $Ca_{in}$ and the dimensionless flow-wise position $x^* = x/L$. The contact angle is assumed to be $\pi/2$, so we take $f(\theta_c)=1$. In these plots, $x^*=0$ and $x^*=1$ represent the inlet and outlet, respectively. The intersection between a given $\dot{\lambda}$ surface and the horizontal plane $\dot{\lambda} = 0$  indicates a transition in (in)stability. Thus, our classification of instability into three regimes becomes clear. Regime I is to the left of the line of intersection, Regime III is to the right of this line, and Regime II refers to cases in which the line of intersection is crossed only for some range of $Ca_{in}$ values. For example, in a parallel cell (Fig.~\ref{fig:GR_para}), the line of intersection is parallel to the $x^*$ axis, thus Regime II corresponds to one specific value of  $Ca_{in}$. For that value of $Ca_{in}$, the interface is neutrally stable. In a diverging cell (Fig.~\ref{fig:GR_div}) or a converging cell (Fig.~\ref{fig:GR_conv}), on the other hand, the line of intersection is crossed for a \emph{range} for $Ca_{in}$ values, as one goes from the inlet to the outlet, i.e., from $x^*=0$ to $x^*=1$. 

Furthermore, by comparing the growth rates in Figs.~\ref{fig:GR_div} and \ref{fig:GR_conv}, we observe that, for larger $Ca_{in}$, the growth rate in a converging cell could be greater than that in a diverging cell. This observation could explain the recent counterintuitive simulation results reported by \citep{Jackson2017}, in which while the perturbation in a converging Hele-Shaw cell has a larger growth rate than in a parallel cell, the perturbation's growth rate in a diverging cell can be smaller, for large $Ca_{in}$, than a in parallel cell.

\section{Direct numerical simulations}
\label{sec:simulations}

Numerical simulations are carried out using the interFoam solver \citep{SD2009,DAT12} in OpenFOAM{\textsuperscript\textregistered} ver.~6.0 \citep{WTJF98}, an open-source library based on the \emph{finite volume method} (FVM) \citep{MMD2016}. In this analysis, the fluids are considered as incompressible, immiscible and unsteady. As depicted in Fig.~\ref{fig:schematic_figure}, the fluids flow through a Hele-Shaw cell with a depth that varies with $x$. The contact angle between the displacing fluid and the wall is taken to be $\theta_c = \pi/2$ (the interface is flat across the depth). Three types of cells are considered: converging (negative depth gradient, $\alpha<0$), parallel (zero gradient, $\alpha=0$), and diverging (positive depth gradient, $\alpha>0$).

Below, we present numerical ``experiments'' (studies) based on water (fluid 1) injected into mineral oil (fluid 2). The corresponding fluid properties, geometric parameters and flow quantities used for these simulations are summarized in Table~\ref{tab:property}. In simulations, the $U_{in}$ values used are back-calculated from $Ca_{in}$.

\begin{table} [ht]
\caption{Fluid properties, geometric parameters and flow quantities used for the numerical simulations.}
\centering
\begin{tabular}{l l}
 \hline
 \hline
 Property & Value (SI units) \\
 \hline
 $\mu_1$ &  $1.0 \times 10^{-3}$ \\
 \hline
 $\rho_1$ &  $1.0 \times 10^{3}$ \\
 \hline
 $\mu_2$ &   $6.50 \times 10^{-2}$ \\
 \hline
 $\rho_2$ &  $8.30\times 10^{2}$ \\
 \hline
 $\gamma$ &  $2.95 \times 10^{-2}$ \\
 \hline
 $\theta_c$ & $\pi/2$ \\
\hline
\end{tabular}
\hspace{1cm}
\begin{tabular}{l l}
 \hline
 \hline
 Quantity & Value (SI units or --~)  \\
 \hline
 $h_{in}$ &  $1 \times 10^{-3}$ \\
 \hline
 $W$ &  $5 \times 10^{-2}$ \\
 \hline
 $L$ &   $2 \times 10^{-1}$ \\
 \hline
 $\alpha$ &  $0,\pm 5 \times 10^{-4}, \pm 10 \times 10^{-4}$ \\
 \hline
 $Ca_{in}$ &  $0.0067$ to $0.0200$ \\
\hline
\end{tabular}
\label{tab:property}
\end{table}

\subsection{Governing equations}

In our DNS study, we solved the ``full'' 3D Navier--Stokes equation directly in each fluid, instead of the simplified and depth-averaged equations commonly solved in previous numerical studies. Specifically, the governing equations solved by the interFoam implementation \citep{DAT12} in OpenFOAM{\textsuperscript\textregistered} are the conservation of mass and momentum for a \emph{two-fluid mixture}, written as:
\begin{align}
\nabla\cdot{\bm{v}} &=0, \label{eq:mass}\\
\frac{\partial (\varrho \bm{v})}{\partial t} + \nabla \cdot (\varrho \bm{v} \otimes \bm{v}) &= -\nabla p +  \nabla \cdot [\eta (\nabla \bm{v} + \nabla \bm{v}^\top )] + \bm{F}, \label{eq:momentum}
\end{align}
where $\varrho=\varrho(x,y,z,t)$ is the mixture density, $\bm{v}=\bm{v}(x,y,z,t)$ is the mixture velocity, $p=p(x,y,z,t)$ is the pressure, $\eta=\eta(x,y,z,t)$ is the mixture viscosity, and $\bm{F}$ is a fictitious body force used to enforce surface tension at the fluid--fluid interface. Physically, this body force due to surface tension results in a pressure jump at the interface, and it is evaluated (per unit volume) by the continuum surface force (CSF) model \citep{BKZ1992,P18}:
\begin{equation}
    \bm{F} = \gamma \kappa \nabla \phi,
\end{equation}
where $\gamma$ is the surface tension, and $\kappa = -\nabla\cdot(\nabla\phi/\|\nabla\phi\|)$ is the mean curvature of the fluid--fluid interface computed directly from its surface normals. Here, $\phi=\phi(x,y,z,t)$ is the phase fraction in a given cell of the computational grid, defined as 
\begin{equation}
\phi   \begin{cases} 
	     \in (0,1) & \Rightarrow\text{Interface;} \\ 
    	 = 1 & \Rightarrow\text{Fluid 1};\\  
   	     = 0 & \Rightarrow\text{Fluid 2}.\\ 
        \end{cases}  
\end{equation}
The phase fraction keeps track of where each fluid (``1'' and ``2'') goes in the computational domain. We stress that this ``mixture'' model does not consider ``different'' physics than the mathematical model in Sect.~\ref{sec:linear}; this is simply a numerical approach to handling dissimilar fluids separated by an immiscible interface. 

In the interFoam solver, the phase fraction $\phi$ is solved using a modified volume-of-fluid (VOF) method \citep{HN81}:
\begin{equation}
\frac{\partial \phi}{\partial t} + \nabla \cdot (\phi \bm{v}) + \nabla \cdot (\phi (1-\phi) \bm{v}_r) =0, 
\label{eq:phi_fract}
\end{equation}
where $\bm{v}_r = \bm{v}_1 - \bm{v}_2$ is the relative velocity between the two fluids. The fluid properties of the mixture, as well as the mixture velocity, to be used in Eqs.~\eqref{eq:mass} and~\eqref{eq:momentum}, can thus be expressed as
\begin{subequations}\begin{align}
\varrho &= \phi \rho_1 + (1-\phi) \rho_2,\\
\eta  &= \phi \mu_1 + (1-\phi) \mu_2,\\
\bm{v} &= \phi \bm{v}_1 + (1-\phi) \bm{v}_2,
\end{align}\label{eq:mu_mix}\end{subequations}
where the subscripts ``$1$'' and ``$2$'' refer to \emph{fluid 1} (displacing) and \emph{fluid 2} (displaced/defending), respectively. Eqs.~\eqref{eq:mass},~\eqref{eq:momentum} and~\eqref{eq:phi_fract} are discretized spatially using the FVM and integrated in time via an Euler implicit scheme. Convection terms are discretized using a linear upwind scheme, while diffusion terms are discretized using a linear scheme. Gauss integration is employed for both terms, and gradients are corrected to account for non-orthogonal fluxes, which occur in angled cells. The Navier--Stokes solution algorithm used in interFoam is ``PIMPLE,'' which is a combination of the \emph{pressure-implicit with splitting of operators} (PISO) method \citep{Issa86,FP_chapter} and the \emph{semi-implicit method for pressure-linked equations} (SIMPLE) \citep{PS72,FP_chapter}. In PIMPLE, PISO is used as the inner loop corrector to update pressure and velocity, and SIMPLE is used as the outer loop corrector to ensure convergence. Thus, PIMPLE achieves a more robust pressure-velocity coupling.

\subsection{Mesh generation}
When performing numerical simulations, on one hand, it is important to have a mesh that is fine enough to capture the physical properties accurately; on the other hand, a mesh with fewer cells saves computational resources. To balance accuracy and efficiency, we only refine the mesh in the region of interest, which in the present problem is the fluid--fluid interface. We use a relatively coarse mesh for the remainder of the Hele-Shaw cell. Since the interface is moving, adaptive mesh refinement is employed. At every time step the mesh is dynamically refined in the spatial region where $0.001< \phi(x,y,z,t) < 0.999$, i.e., close to the interface to ensure it is well resolved. The initial mesh resolution is given in Table~\ref{tab:GridIndTest}.

\begin{table}[h]
\caption{Mesh generation for grid and time independence test(s) parameters for a parallel cell. For converging and diverging cells, only $\Delta z$ changes.}
\centering
\begin{tabular}{l l l l}
 \hline
 \hline
 Case ID & 1 &  2 &  3 \\
 \hline
 Grid elements & $750$  & 6,000  & 48,000 \\
 Grid resolution, $\Delta x$ [m] & $8\times 10^{-3}$  & $4\times 10^{-3}$  & $2\times 10^{-3}$ \\
 Grid resolution, $\Delta y$ [m] & $5\times 10^{-3}$  & $2.5\times 10^{-3}$  & $1.25\times 10^{-3}$ \\
 Grid resolution, $\Delta z$ [m] & $3.33 \times 10^{-4}$  & $1.67 \times 10^{-4}$  & $8.33 \times10^{-5}$ \\ 
 \hline
 Time step [s] & $10^{-3}$  & $10^{-4}$  & $10^{-5}$  \\
\hline
\label{tab:GridIndTest}
\end{tabular}
\end{table}

\subsection{Initial and boundary conditions}
\begin{figure}[t]
  \centering
    \begin{subfigure}{0.75\textwidth}
    \includegraphics[width=\textwidth]{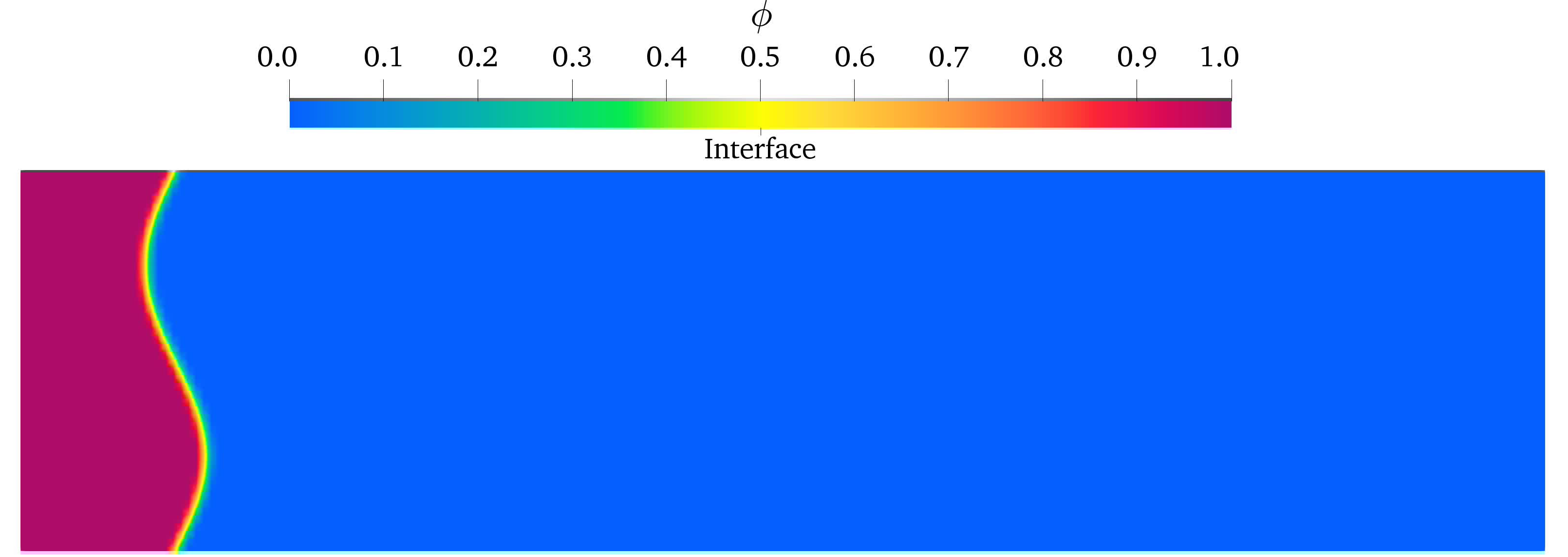}
    \subcaption{initial interface perturbation}
    \label{fig:interface_example_initial}
  \end{subfigure}
  \hfill
  \begin{subfigure}{0.75\textwidth}
    \includegraphics[width=\textwidth]{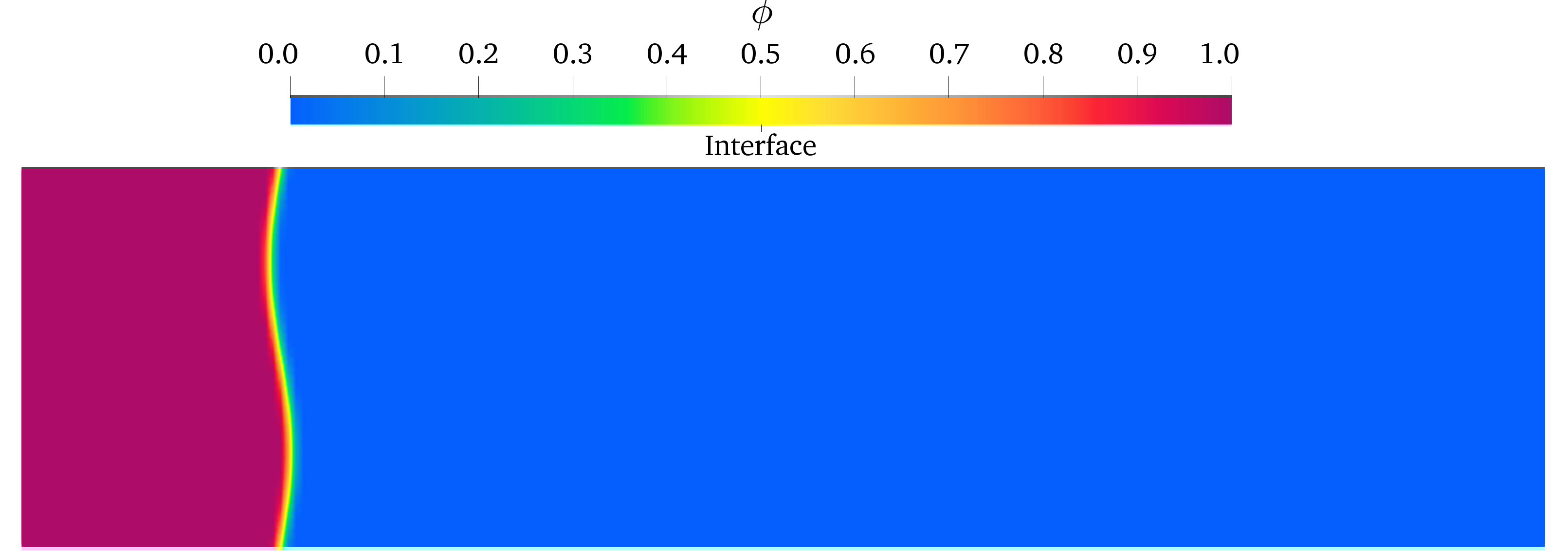}
    \subcaption{interface evolution, stable case: $Ca_{in}=0.01$, $t=45s$}
    \label{fig:interface_example_stable}
  \end{subfigure}
   \begin{subfigure}{0.75\textwidth} 
    \includegraphics[width=\textwidth]{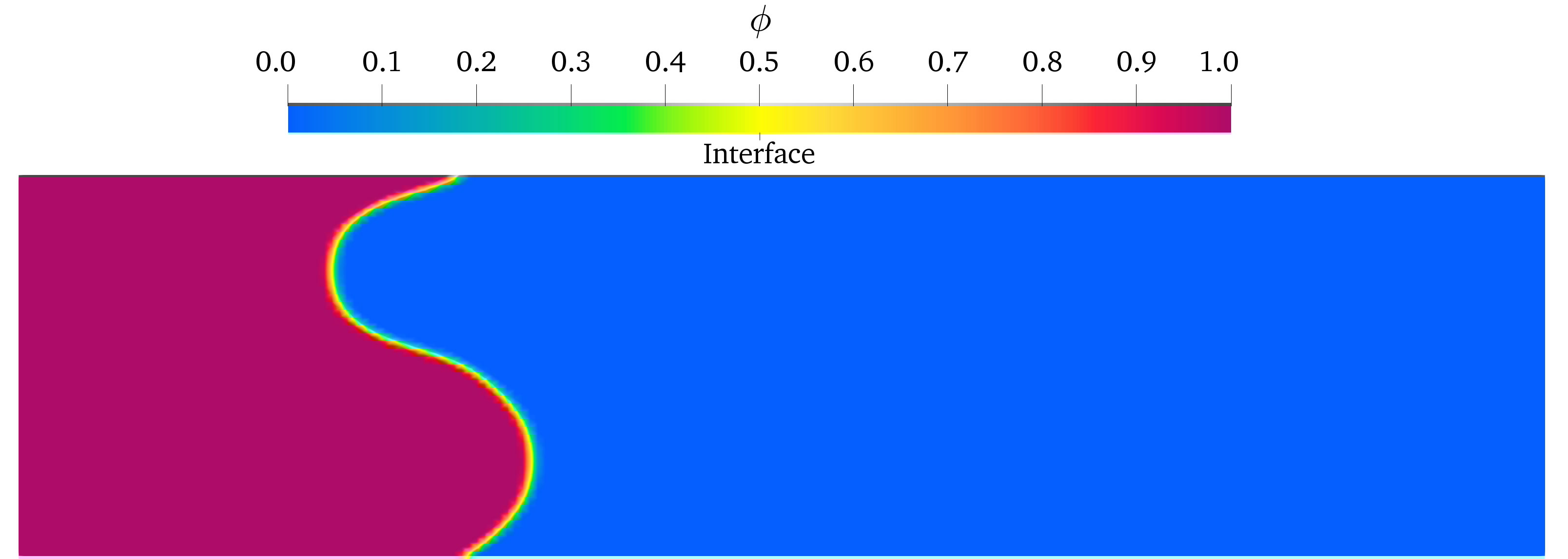}
    \subcaption{interface evolution, unstable case: $Ca_{in}=0.03$, $t=45s$}
    \label{fig:interface_example_unstable}
  \end{subfigure}
\caption{Top view of example simulation of interface evolution in a converging cell ($\alpha = -5 \times 10^{-4}$).}
\label{fig:interface_example}
\end{figure}

At the inlet ($x=0$), we employ a horizontal velocity profile that satisfies no-slip at $z=0,h_{in}$, i.e., $u_x(z) = -6 U_{in}(z/h_{in}-1/2)^2 + (3/2) U_{in}$, as the boundary condition for the mixture velocity. A zero-gradient boundary condition is employed for the flow at the outlet ($x=L$), except the pressure, which is fixed to zero to set the gauge. Initially, fluid 1 and fluid 2 are separated by a flat interface $x = \zeta_0= 20 ~\mathrm{mm}$. An initial perturbation is applied at the interface, i.e., $\zeta(y,0) = \zeta_0 [1+\epsilon \sin (ky)]$ (see Fig.~\ref{fig:interface_example_initial}). The initial magnitude of the perturbation is set by $\epsilon=0.2$, and its wavenumber is $k = 2 \pi /W$. Along the top ($z=h(x)$) and bottom ($z=0$) plates of the cell, a no-slip boundary condition is prescribed. All variables are assumed to be periodic at the side (lateral) ends, $y=0$ and $y=W$, of the cell. An example simulation under these initial and boundary conditions is shown in Fig.~\ref{fig:interface_example}, highlighting how the initial perturbation can grow or decay, and how the interface remains sharp due to adaptive mesh refinement.

\subsection{Grid and time step independence test}
Three sets of simulations on different meshes, as listed in Table~\ref{tab:GridIndTest}, were conducted to show grid and time step independence. We compute the deviation of the finger's length, defined as $\xi(t) := \max_y[\zeta(y,t)] - \zeta_0(t)$, and found from linear stability analysis to be $\xi_{\mathrm{la}}(t)$, to the simulation prediction  $\xi_{\mathrm{ns}}(t)$, as:
\begin{equation}
    \varepsilon(t) = \left|  \frac{\xi_{\mathrm{ns}}(t) - \xi_{\mathrm{la}}(t)}{\xi_{\mathrm{la}}(t)} \right|.
\label{eq:error_theor_analytical}
\end{equation}

\begin{figure}[t]
  \centering
  \begin{subfigure}{0.32\textwidth}
    \includegraphics[width=\textwidth]{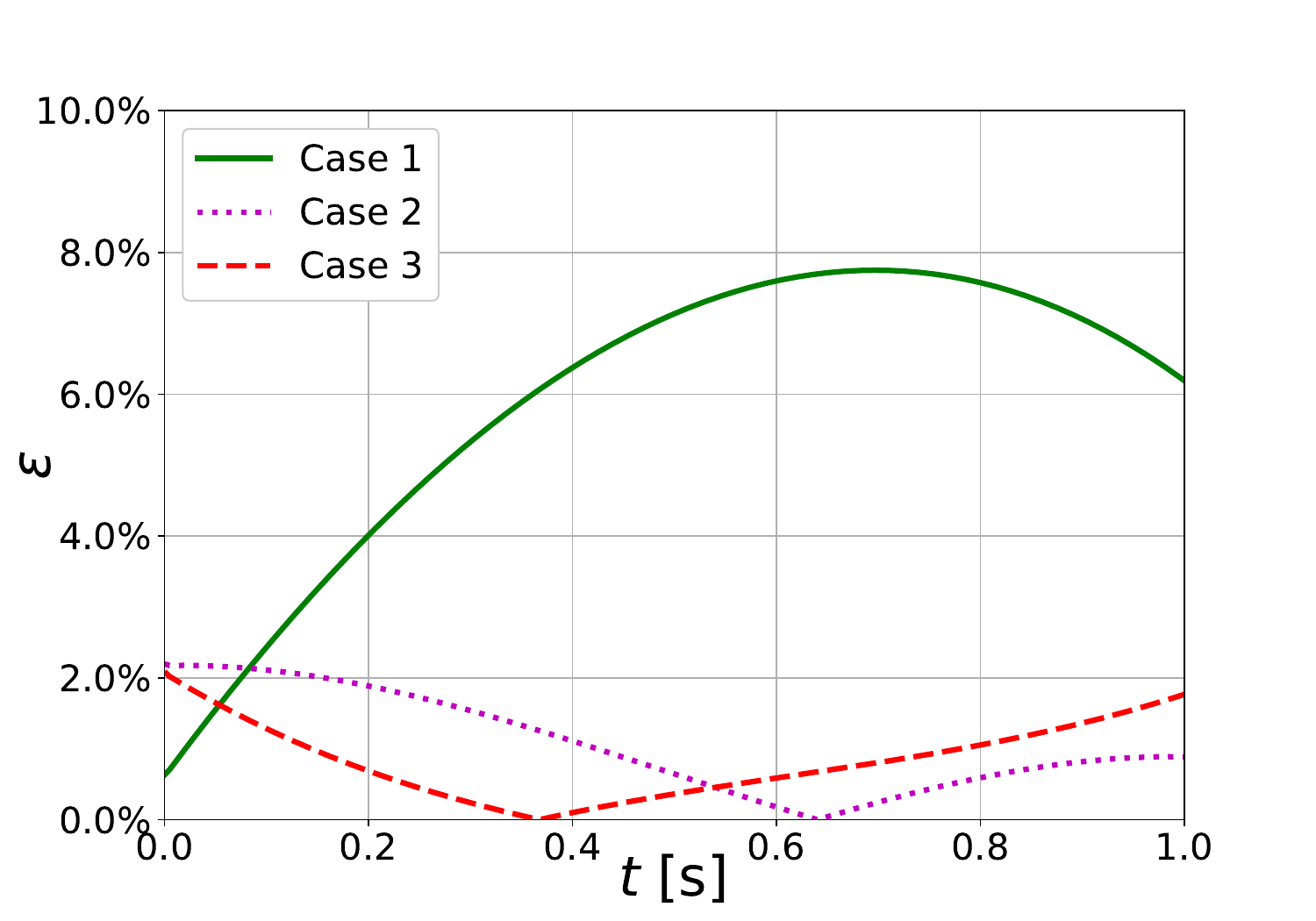}
    \subcaption{parallel cell ($\alpha = 0$)}
  \end{subfigure}
  \hfill
  \begin{subfigure}{0.32\textwidth}
    \includegraphics[width=\textwidth]{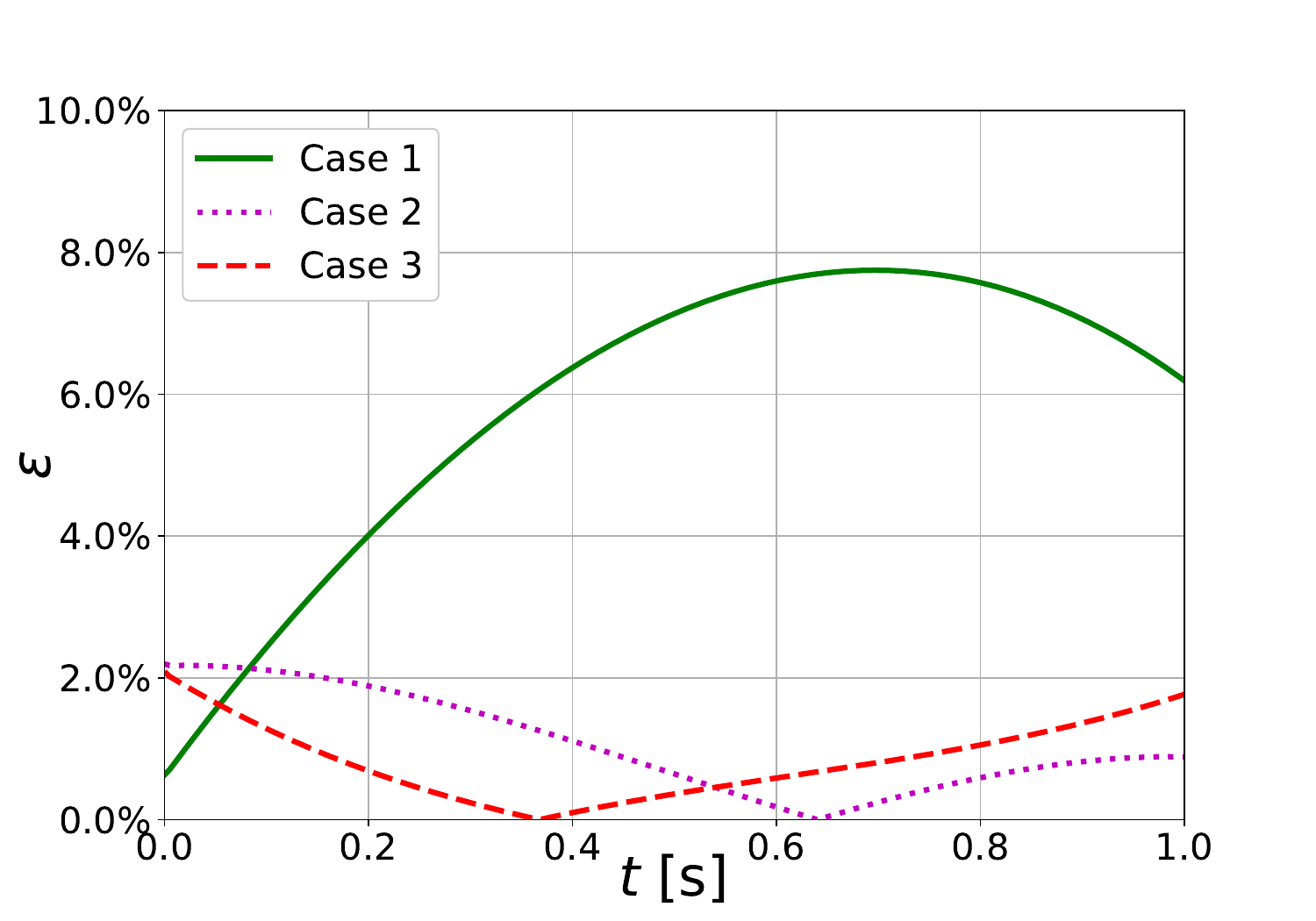}
    \subcaption{diverging cell ($\alpha = 5 \times 10^{-4}$)}
  \end{subfigure}
  \hfill
  \begin{subfigure}{0.32\textwidth} 
    \includegraphics[width=\textwidth]{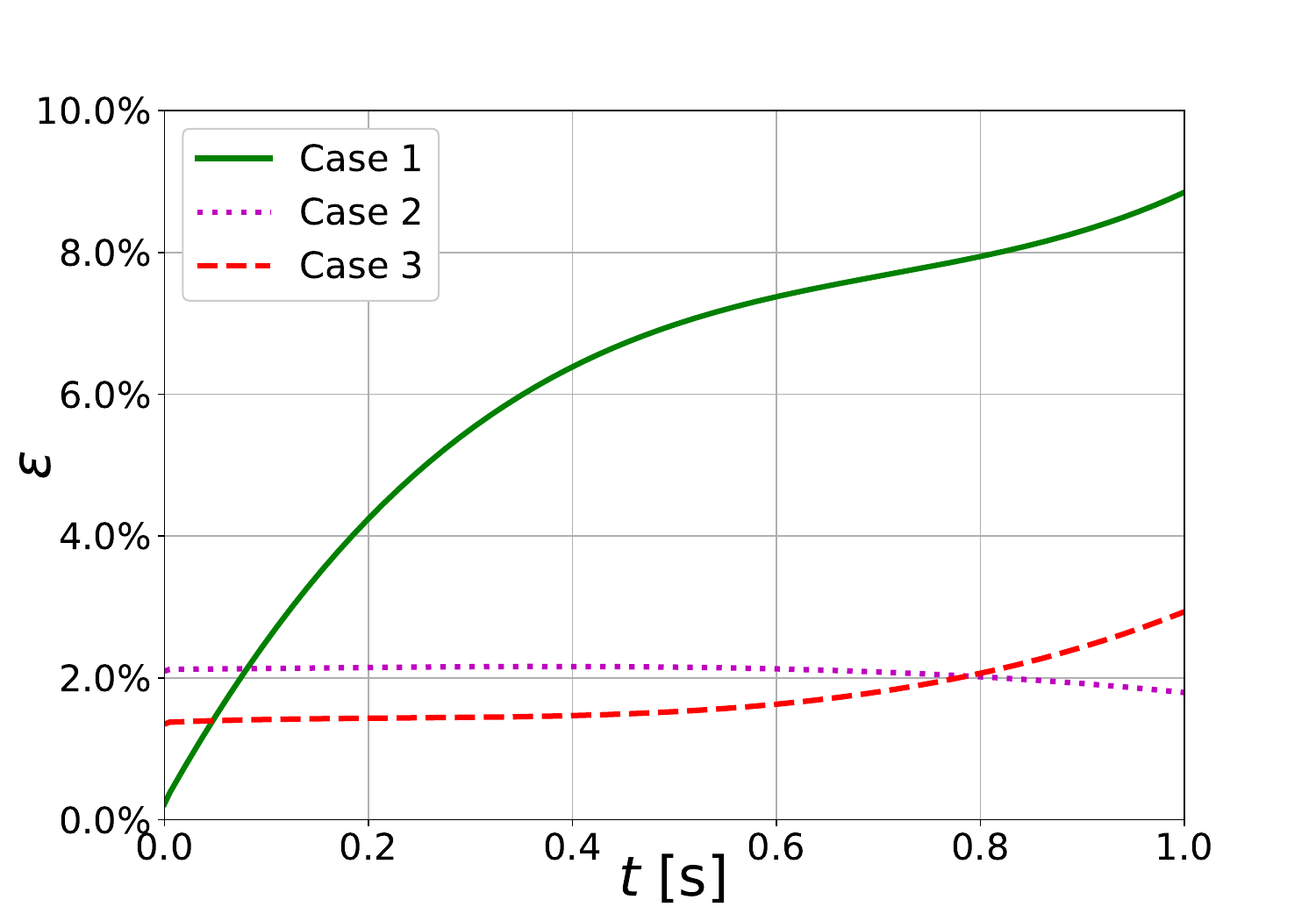}
    \subcaption{converging cell ($\alpha = -5 \times 10^{-4}$)}
  \end{subfigure}
\caption{Grid and time step independence tests show the \% difference in the fingers length (between linear stability and numerical simulation) as a function of time. Clearly, the simulation parameters in case 2 and case 3 lead to very similar results, so case 2 shall be employed for all simulations herein.}
\label{fig:ind}
\end{figure}

The results in Fig.~\ref{fig:ind} show that the difference in the values of $\varepsilon(t)$ between case 2 and case 3 (less than $1\%$) is insignificant for the purposes of this study. Therefore, for the remainder of the present work, we will employ case 2 as the simulation grid of choice. This grid is less demanding in terms of computational resources than the grid from case 3.

\section{Results: Comparing theory to simulations and stability regimes}
\label{sec:results}

\subsection{Verification of the linear stability analysis}
\label{sec:verification}

\citet{hws16} already performed extensive verification of the interFoam solver's use for simulating viscous fingering in a Hele-Shaw cell of constant depth. Importantly, they showed that simulations can capture quite accurately experimental and theoretical predictions about the length and width of a \emph{single} finger, as in the classical \citet{Saffman1958} experiments and analysis.

In this subsection, we explore flows in angled Hele-Shaw cells with different capillary numbers for converging ($\alpha < 0$), parallel ($\alpha = 0$) and diverging ($\alpha > 0$) cell geometries as cataloged by the `cases' in Table~\ref{tab:RegimeVal}. We wish to verify the linear stability theory from Sec.~\ref{sec:linear} through numerical simulations. Before we discuss the numerical results in detail, it is instructive to make a few remarks. First, the linear analysis holds, strictly speaking, only at the moment of initiation (onset) of instability. Nevertheless, the linear analysis is generally used in the literature to describe the unstable interface's evolution. Additionally, here for simplicity and for consistency with some previous analyses, only one Fourier mode was used to represent the perturbation. At later times in the displacement process, the simulations capture the full nonlinear unstable interface evolution, which is expected to be qualitatively similar (but quantitatively different) than the prediction of the linear analysis. This specific point is of interest to us in the present work. Even when linear theory does not provide precise quantitative prediction about the interface's evolution (and, instead, direct numerical simulations must be used), linear theory correctly delineates the stability regimes and their transitions in the capillary number--depth gradient, i.e., $(Ca,\alpha)$, space.

\begin{figure}
  \centering
  \begin{subfigure}[b]{0.93\textwidth}
    \includegraphics[width=\textwidth]{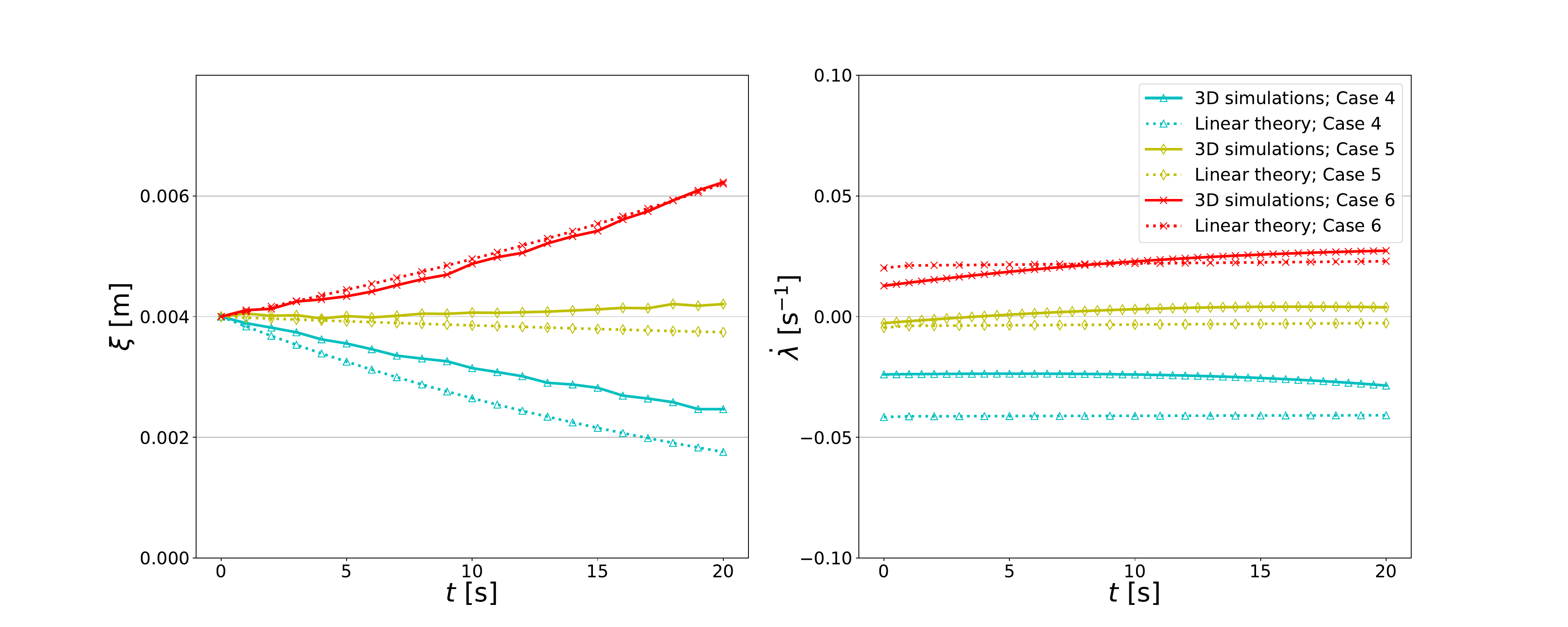}
    \subcaption{a converging cell with $\alpha = -5 \times 10^{-4}$}
    \label{fig:val_conv}
  \end{subfigure}
      \begin{subfigure}[b]{0.93\textwidth}
    \includegraphics[width=\textwidth]{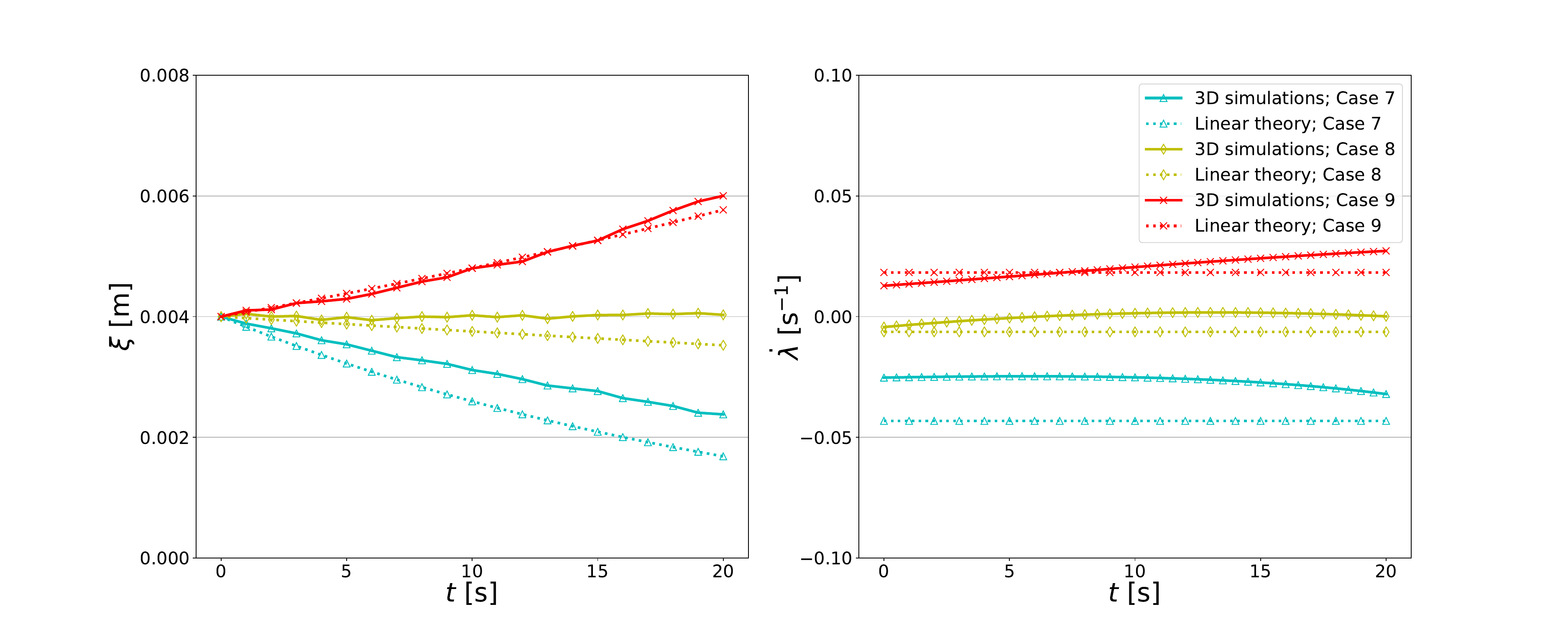}
    \subcaption{a parallel cell with $\alpha = 0$}
    \label{fig:val_para}
  \end{subfigure}
\begin{subfigure}[b]{0.93\textwidth} 
    \includegraphics[width=\textwidth]{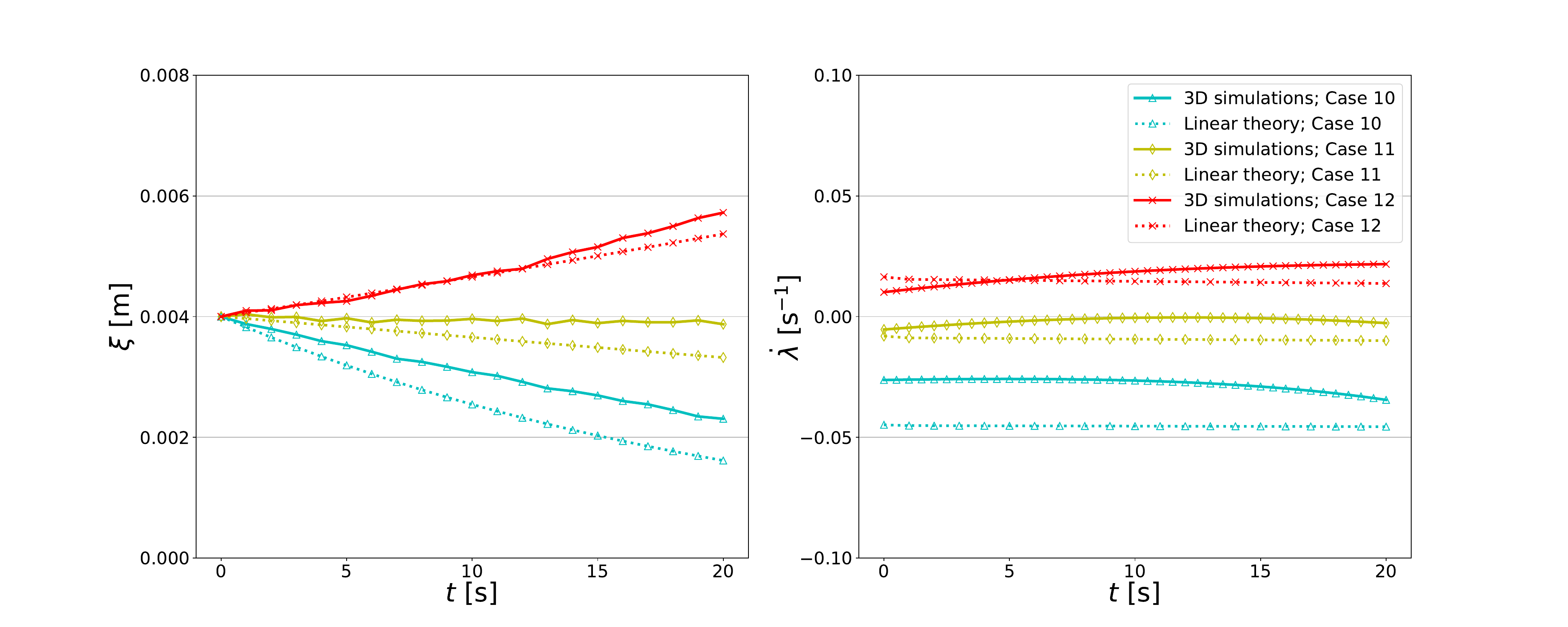}
    \subcaption{a diverging cell with $\alpha = 5 \times 10^{-4}$}
    \label{fig:val_div}
  \end{subfigure}
\caption{Linear stability analysis verification for the finger's length $\xi(t)$ and the growth rate $\dot{\lambda}(t)$ in (a) a converging cell (Cases 4, 5, 6), (b) a parallel cell (Cases 7, 8, 9), (c) a diverging cell (Cases 10, 11, 12). Colors represent different inlet $Ca$, as labelled. Simulations results (solid curves) verify the three regimes theory: blue solid curves (decreasing $\xi$ and $\dot{\lambda}<0$) are in Regime I, yellow solid curves (roughly constant $\xi$ and $\dot{\lambda}\approx 0$) are in Regime II, and red solid curves (increasing $\xi$ and $\dot{\lambda}>0$) are in Regime III.}
\label{fig:val}
\end{figure}

The first verification results are presented in Fig.~\ref{fig:val} for each of the three types of Hele-Shaw cells in panels (a), (b) and (c). Solid and dotted lines refer to numerical and theoretical results, respectively. Lines with different symbols represent different cases from Table~\ref{tab:RegimeVal} with different inlet capillary numbers $Ca_{in}$. The finger's length $\xi$ (left panels) and growth rate $\dot{\lambda}(t)$ (right panels) are plotted as functions of time. The theoretical predictions are computed from Eqs.~\eqref{eq:iface-perturb} and \eqref{eq:GR} with $\xi(t) = \max_y[\zeta(y,t)] - \zeta_0(t)$. Numerical results are post-processed from the simulation data; the finger's length is calculated as $\xi(t) = \{ \max_y[\zeta(y,t)] + \min_y[\zeta(y,t) \}/2$, and the growth rate is calculated as $\dot{\lambda}(t) = [1/\xi(t)]d\xi/dt$ using the cubic-spline interpolation with smoothing available in SciPy \citep{SciPy}.

\begin{table}[ht]
\caption{Classification of cases for simulations used for verification of the linear theory.}
\centering
\begin{tabular}{l l l l}
 \hline
 \hline
   & {$Ca_{in} = 0.0067$} & {$Ca_{in} = 0.0147$} & {$Ca_{in} = 0.0200 $} \\
 \hline
 Converging cell ($\alpha = -5 \times 10^{-4}$) & Case 4  &  Case 5 &  Case 6 \\
 \hline
\multicolumn{1}{l}{Parallel cell ($\alpha = 0$)}& Case 7  &  Case 8 &  Case 9 \\
\hline
\multicolumn{1}{l}{Diverging cell ($\alpha = 5 \times 10^{-4}$)} & Case 10  &  Case 11 &  Case 12 \\
\hline
\label{tab:RegimeVal}
\end{tabular}
\end{table}

Numerical results and linear stability analysis agree very well at {$Ca_{in} = 0.0200$} in each cell: Case {6} in Fig.~\ref{fig:val_conv}; Case {9} in Fig.~\ref{fig:val_para}; and Case {12} in Fig.~\ref{fig:val_div}. In these cases, the finger's length $\xi(t)$ {increases} over time and the growth rate $\dot{\lambda}(t)$ remains {positive}, meaning that the interface is {unstable}, which corresponds to {Regime III} in classification introduced in Sec.~\ref{sec:linear}. At {$Ca_{in} = 0.00147$}, numerical simulations show a roughly constant finger's length. The growth rate is almost zero (Case 8 in Fig.~\ref{fig:val_para}), or suffers a change of sign (Case 5 in Fig.~\ref{fig:val_conv} and Case 11 in Fig.~\ref{fig:val_div}). At this value of $Ca_{in}$ we are reminded of Regime II from the linear stability analysis. As $Ca_{in}$ is {decreased} to {$0.0067$}, simulations show that $\xi(t)$ continues to {decay}, and $\dot{\lambda}(t)$ is always {negative}: Case {4} in Fig.~\ref{fig:val_conv}; Case {7} in Fig.~\ref{fig:val_para}; and Case {10} in Fig.~\ref{fig:val_div}.

Therefore, the prediction from linear stability theory regarding three instability regimes has been verified through numerical simulations. However, the quantitative prediction of the growth rate from linear analysis is most accurate in Regime III, which corresponds to the classical Saffman--Taylor instability. In Regime I and II, our results quantify the difference between the linear model and 3D DNS. Indeed, as the interface becomes flattened, it is increasingly less meaningful to try to define the finger's length as the difference between maximum and minimum extent. In this case, integrated metrics such as the isoperimetric ratio have been preferred by other authors \citep{MMM19}. Moreover, we have verified that the evolution of the perturbations according to linear stability theory and DNS match more closely (quantitatively) for smaller perturbation magnitudes, e.g., taking $\epsilon = 0.05$ or even $0.02$. However, to better resolve these very small (and slower growing perturbations), a finer mesh should be used. The effect of the perturbation magnitude $\epsilon$ is not an emphasis of the present study, hence these details are omitted for brevity. As mentioned earlier, we use the small but not infinitesimal value of $\epsilon = 0.2$ in all of our DNS results.

\subsection{The effect of the flow-wise depth gradient}
\label{sec:gradient}

\begin{figure}
  \centering
    \begin{subfigure}[b]{0.58\textwidth}
    \includegraphics[width=\textwidth]{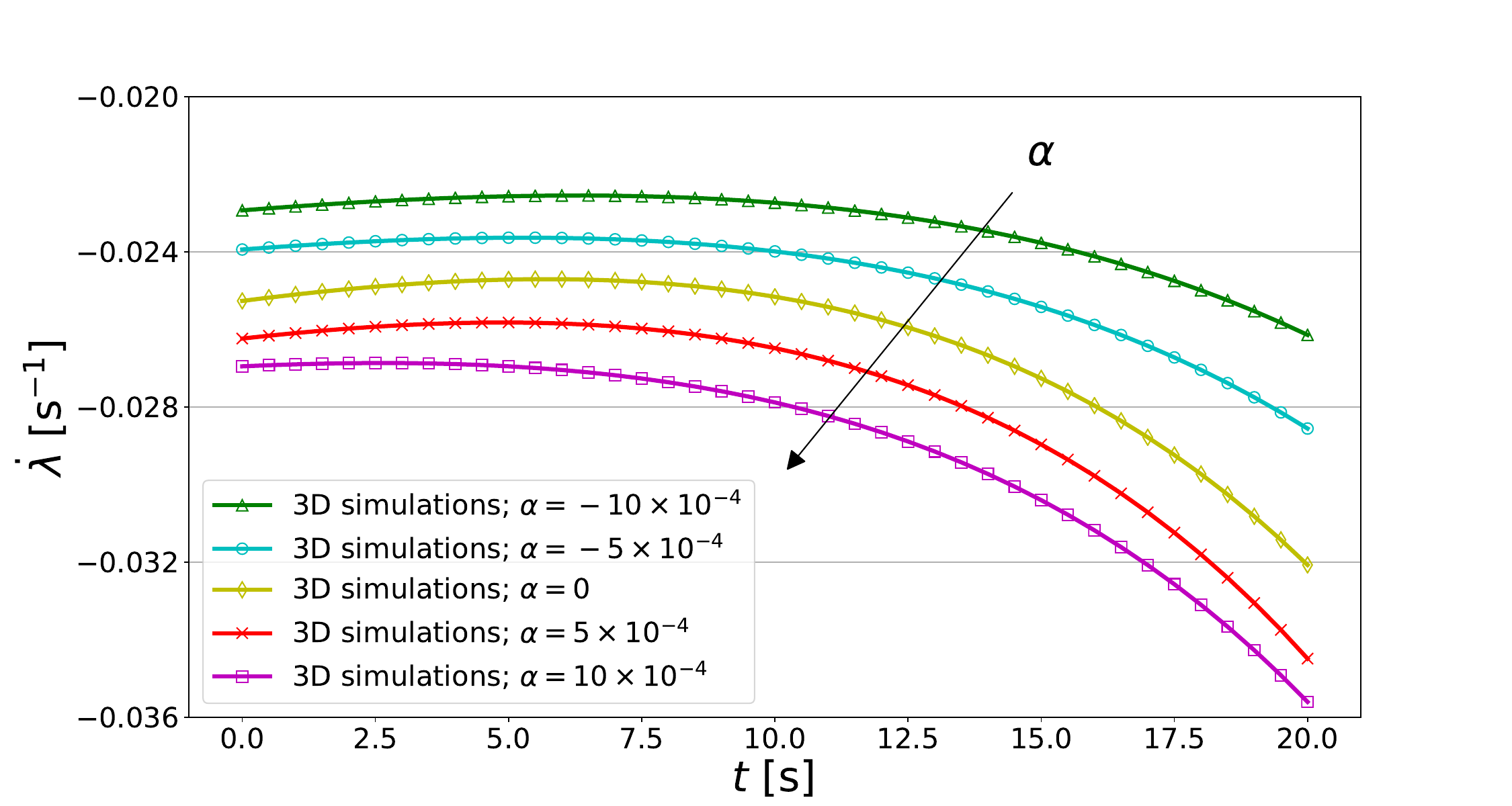}
    \subcaption{Regime I ($Ca = 0.0067$)}
    \label{fig:growth_rate_regimeI}
  \end{subfigure}
    \begin{subfigure}[b]{0.4\textwidth}
    \includegraphics[width=\textwidth]{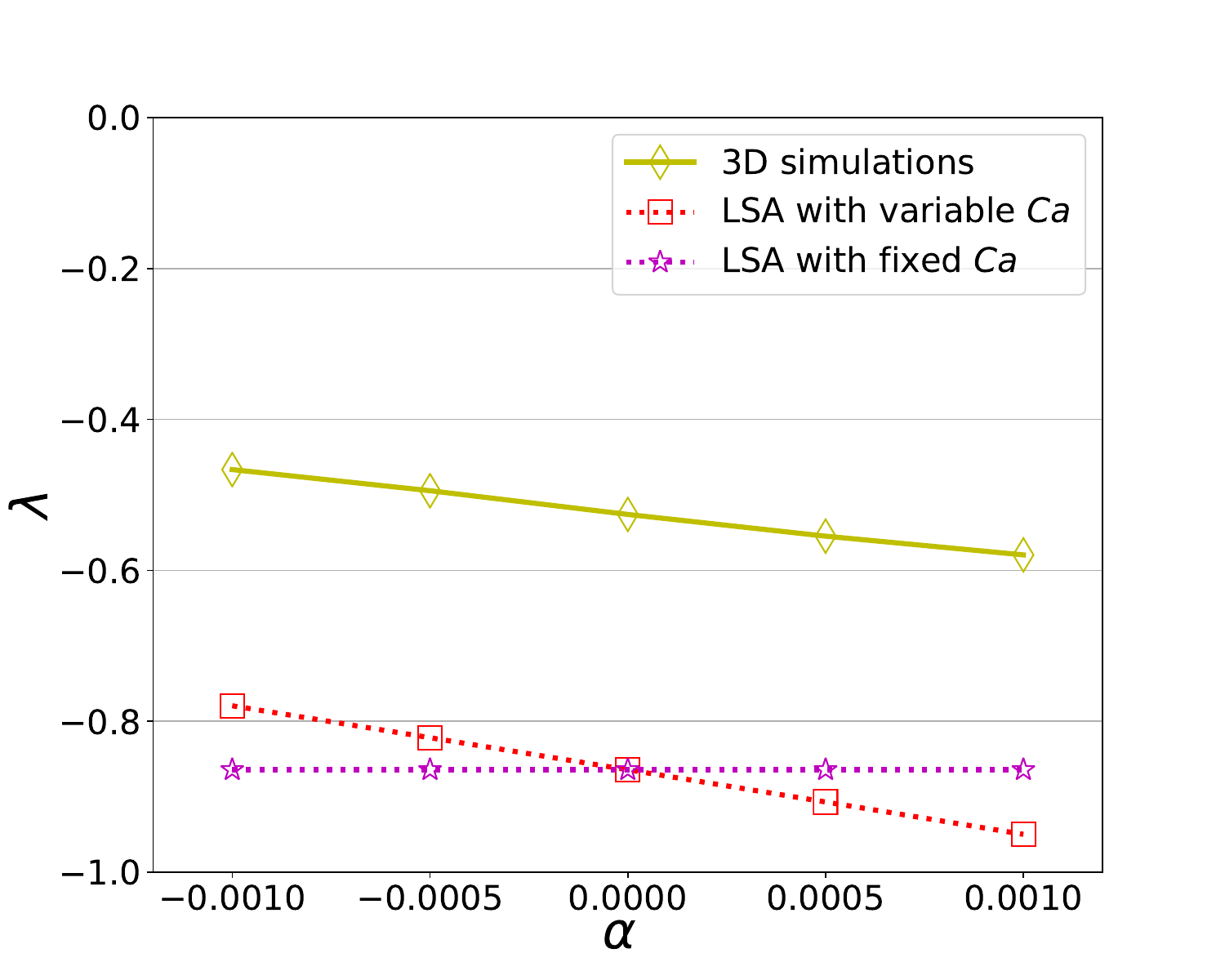}
    \subcaption{Regime I ($Ca = 0.0067$) }
    \label{fig:val2_regime1}
  \end{subfigure}
  \begin{subfigure}[b]{0.58\textwidth}
    \includegraphics[width=\textwidth]{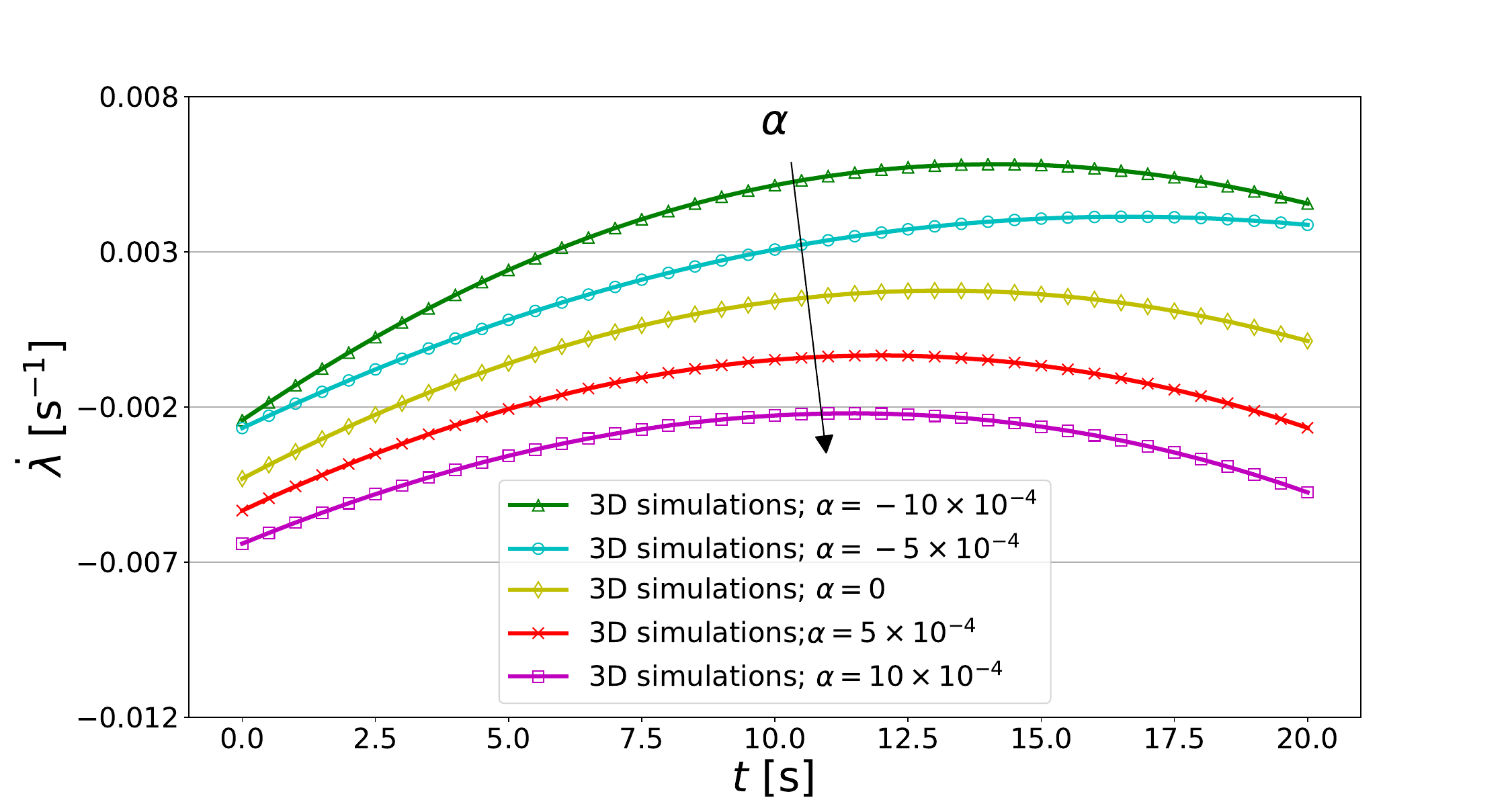}
    \subcaption{Regime II ($Ca = 0.0147$)}
    \label{fig:growth_rate_regimeII}
  \end{subfigure}
\begin{subfigure}[b]{0.4\textwidth}
    \includegraphics[width=\textwidth]{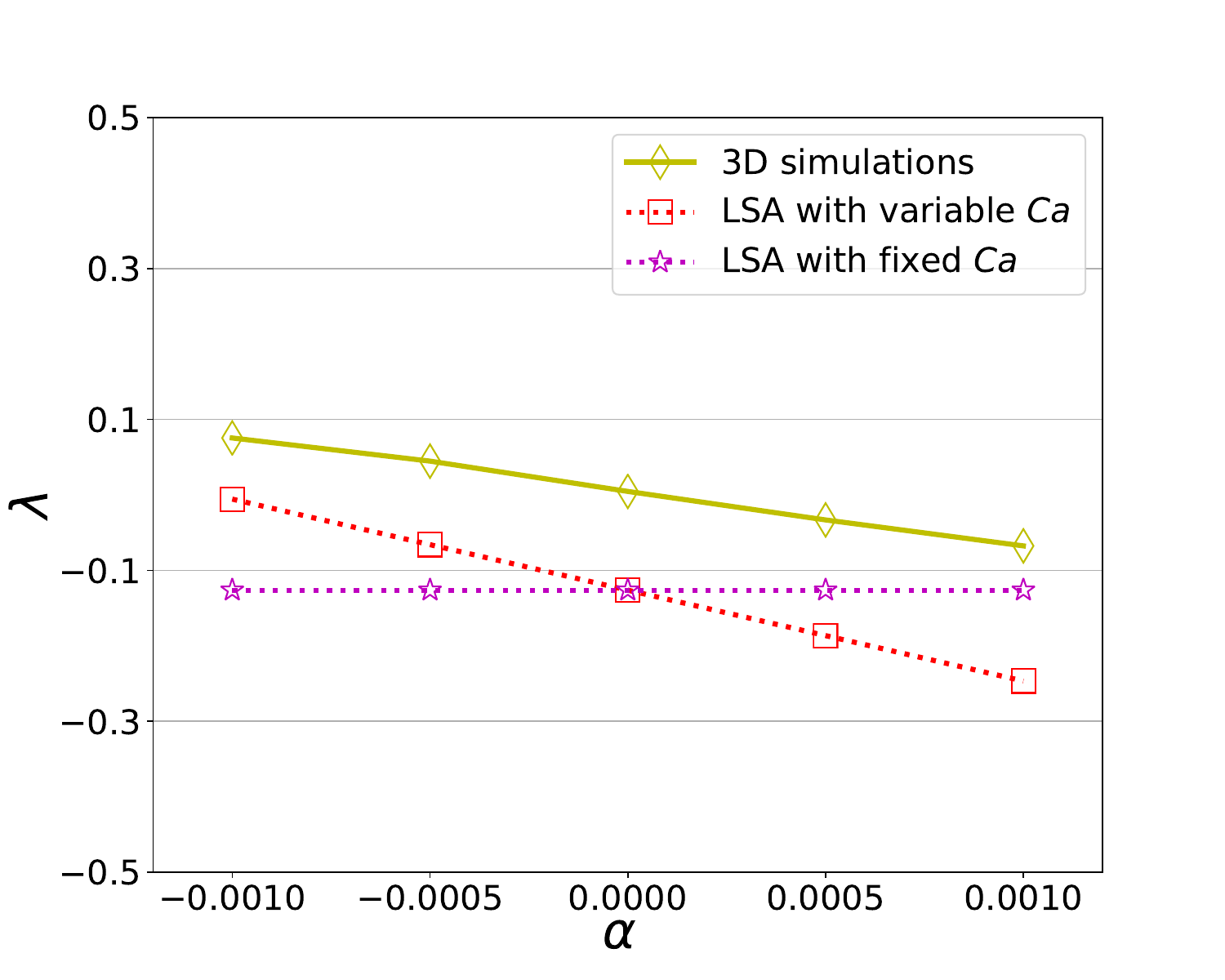}
    \subcaption{Regime II ($Ca = 0.0147$) }
    \label{fig:val2_regime2}
  \end{subfigure}
\begin{subfigure}[b]{0.58\textwidth} 
    \includegraphics[width=\textwidth]{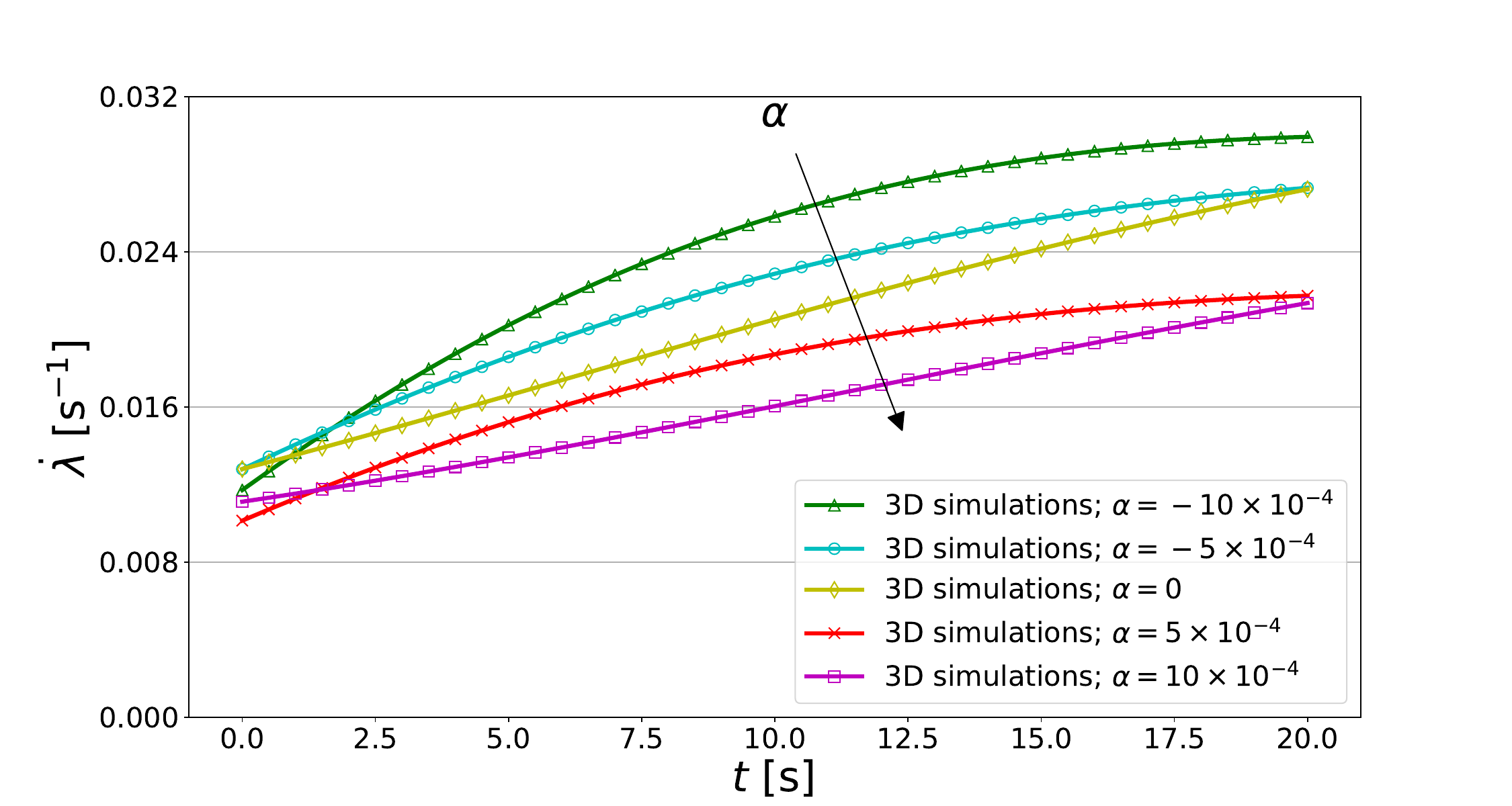}
    \subcaption{Regime III ($Ca = 0.0200$)}
    \label{fig:growth_rate_regimeIII}
  \end{subfigure}
  \begin{subfigure}[b]{0.4\textwidth} 
    \includegraphics[width=\textwidth]{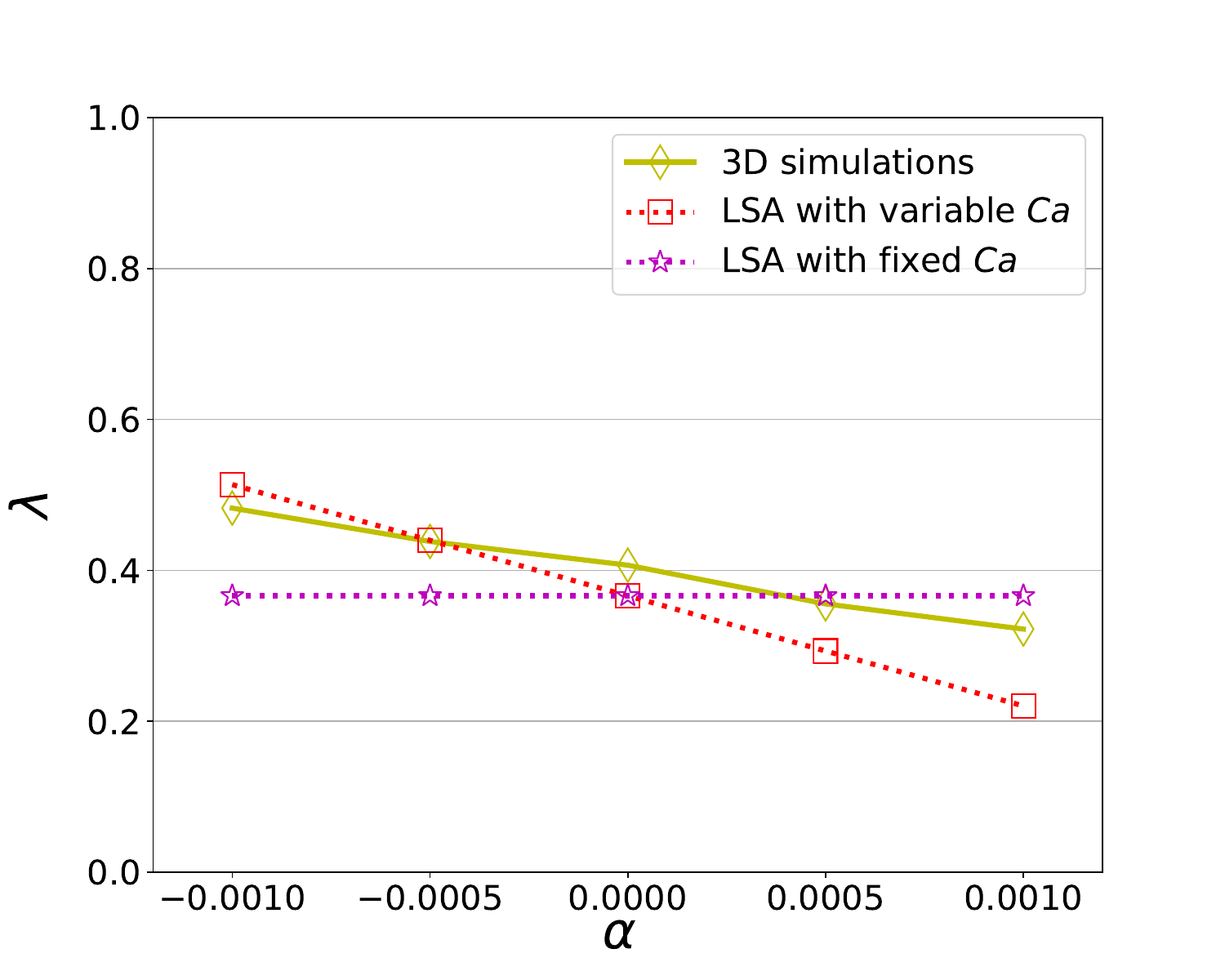}
    \subcaption{Regime III ($Ca = 0.0200$) }
    \label{fig:val2_regime3}
  \end{subfigure}
\caption{The effect of depth gradient $\alpha$ on the growth rate $\dot{\lambda}(t)$ (a,c,e) and the integral of the growth rate $\lambda$ (b,d,f), in the three regimes. As $\alpha$ increases, $\dot{\lambda}$ decreases. The growth rate from 3D simulations is not constant but varies with time, which qualitatively verifies our linear stability analysis from  Sec.~\ref{sec:linear}, denoted as ``LSA with variable $Ca$.'' ``LSA with fixed $Ca$'' stands for the quasi-steady analysis of \citet{Al-Housseiny2012}. LSA with fixed $Ca$ does not capture the decreasing trend of $\lambda$ with $\alpha$ obtained by the present LSA with variable $Ca$ and 3D DNS. In turn, the latter exaggerates the gradient's effect compared to DNS.}
\label{fig:growth_rate}
\end{figure}

In this subsection, we investigate the effect of the depth gradient $\alpha$ on the growth/decay of perturbations on the fluid--fluid interface. The various cases with different values of $\alpha$ are cataloged in Table~\ref{tab:gradient}. The theoretical predictions and numerical results for the growth rate $\dot{\lambda}(t)$, as defined in Sec.~\ref{sec:linear}, in these different (in)stability regimes are shown in Figs.~\ref{fig:growth_rate_regimeI}, \ref{fig:growth_rate_regimeII} and \ref{fig:growth_rate_regimeIII}, respectively, for different values of $\alpha$. The results show that as $\alpha$ increases, $\dot{\lambda}$ decreases. In other words, following the terminology most recently used in \citep{ADM18}, the diverging cells have a relatively stabilizing effect in all three regimes, while converging cells relatively destabilize the interface, compared to the interface evolution in a parallel cell for the same value of $Ca_{in}$. Specifically, the growth rate from 3D simulations is not constant, but varies with time, which qualitatively verifies our novel extension of previous linear stability analysis (Sec.~\ref{sec:linear}).

\begin{table}
\caption{Classification of the simulations conducted to ascertain the effect of the depth gradient.}
\centering
\begin{tabular}{l l l l}
 \hline
 \hline
   & Regime I  &  Regime II &  Regime III \\
   & {$Ca_{in} = 0.0067$} & {$Ca_{in} = 0.0147$} & {$Ca_{in} = 0.0200 $} \\
 \hline
\multicolumn{1}{l}{$\alpha = -10 \times 10^{-4}$} & Case 13  &  Case 14 &  Case 15 \\
  \hline
\multicolumn{1}{l}{$\alpha = -5 \times 10^{-4}$} & Case 4  &  Case 5 &  Case 6 \\
 \hline
\multicolumn{1}{l}{$\alpha = 0$}& Case 7  &  Case 8 &  Case 9 \\
\hline
\multicolumn{1}{l}{$\alpha = 5 \times 10^{-4}$} & Case 10  &  Case 11 &  Case 12 \\
\hline
\multicolumn{1}{l}{$\alpha = 10 \times 10^{-4}$} & Case 16  &  Case 17 &  Case 18 \\
\hline
\label{tab:gradient}
\end{tabular}
\end{table}

To further compare the effect of the depth gradient in each regime, in Fig.~\ref{fig:val2_regime1}, \ref{fig:val2_regime2} and \ref{fig:val2_regime3}, we present the integral of growth rate over time, i.e., $\lambda(t) = \int_0^t \dot{\lambda}(t')\,dt' = \int_{\xi(0)}^{\xi(t)} d\xi'/\xi' = \ln|\xi(t)| - \ln|\xi(0)|$, which is computed over the first $20~\mathrm{s}$ of simulation time. In Regime I (see Fig.~\ref{fig:growth_rate} top row), the simulation results show that the depth gradient has a slight effect on the interface: the integral of the growth rate $\lambda$ decreases slightly with $\alpha$. The present linear stability analysis exaggerates the gradient's effect: the predicted slope of $\lambda$, as a function of $\alpha$, is larger than the one from DNS. We conjecture that a weakly-nonlinear stability analysis (e.g., extending the work of \citep{Miranda1998} to the case of an angled Hele-Shaw cell), which keeps higher-order terms in the perturbation expansion, could correct this exaggeration. Moreover, since $\lambda$ is decreasing and negative, then the suppression of viscous fingering that exists in Regime I is most effective in \emph{diverging} cells, and the larger $\alpha$, the better. This result is somewhat counterintuitive compared to discussion in \citep{Al-Housseiny2012} wherein converging cells are described as the most stabilizing; however, in \citep{Al-Housseiny2012} the three regime diagram proposed herein was obviously not considered.

In Regime II (see Fig.~\ref{fig:growth_rate} middle row), the effect of the gradient is stronger than in Regime I: as $\alpha$ increases, $\lambda$ decreases more rapidly. The gap gradient has the most significant effect in Regime III (see Fig.~\ref{fig:growth_rate} bottom row), indicating that the triggering of fingering in Regime III is most clearly observed in converging cells, and the larger $|\alpha|$, the better. Returning to the comparison with the previous linear stability analysis from \citep{Al-Housseiny2012} (dotted lines with stars in Fig.~\ref{fig:val2_regime1}, \ref{fig:val2_regime2} and \ref{fig:val2_regime3}), we observe that it does not capture the decreasing trend of $\lambda$ with $\alpha$, specifically because in our simulations we have taken a contact angle of $\theta_c = \pi/2$. 

In general, the linear stability analysis provides a theoretical explanation for the DNS results, specifically in the prediction of the dependence of the growth rate on the depth gradient $\alpha$ in an angled Hele-Shaw cell (Fig.~\ref{fig:val2_regime1}, \ref{fig:val2_regime2} and \ref{fig:val2_regime3}). As the capillary number decreases, the discrepancy between theory and simulation increases, as is to be expected for this $Ca\ll1$ linear theory. In particular, one way to understand this observation is to note that as $Ca_{in}$ decreases, the interface becomes more and more flattened, and the finger's length is no longer a sensitive metric.

\subsection{Stability diagram}
\label{sec:sd}

\begin{figure}
\centering
\includegraphics[width=0.7\textwidth]{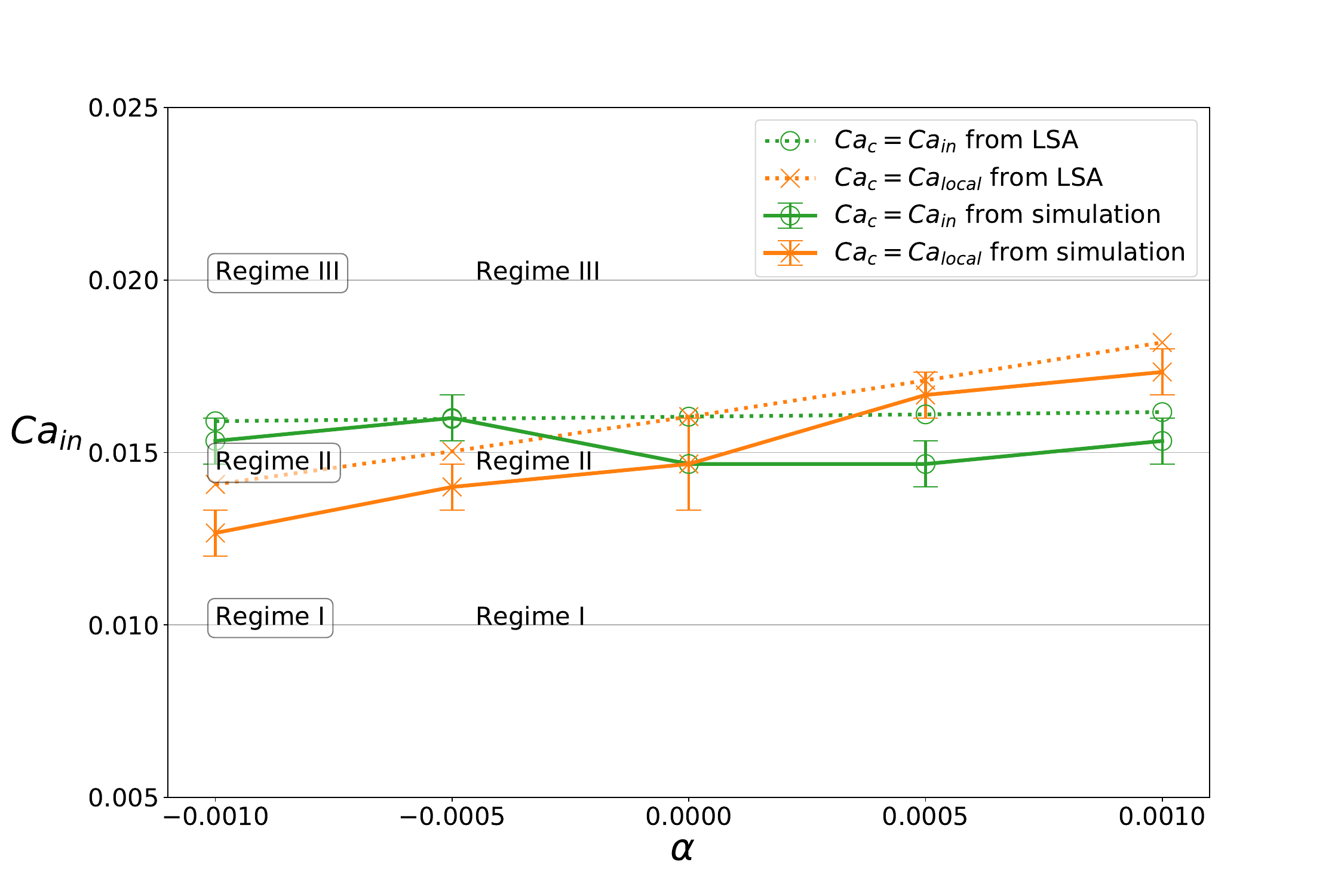}
\caption{Stability Diagram: theoretical and numerical results are plotted as dashed and solid curves, respectively. Green and orange curves represent the critical situations such that $Ca_c = Ca_{in}$ and $Ca_c = Ca_{local}$, respectively. The error bars represent the  difference in $Ca_{in}$ between two simulations. The boxed regimes are those defined by the linear theory and the others are those defined via direct numerical simulations.}
\label{fig:StabilityDiagram}
\end{figure}

In Sec.~\ref{sec:verification}, the linear analysis was verified through numerical simulations of flows throughout the three regimes of the proposed stability classification. However, for marginal cases near the dividing curves between two regimes, we observed that the predictions of the linear theory do not quantitatively agree with DNS. To better understand this discrepancy, we conducted further numerical experiments to explore the numerical regime map for various depth gradient values $\alpha$. In this subsection, we compare the numerical regime diagram to the theoretically predicted one. The result is shown in Fig.~\ref{fig:StabilityDiagram}. Other stability diagrams from previous research \citep{BT18,KXL2017,Brandao2018} have shown the (in)stability by linear stability analysis, experiments, or both. Here, we supplement the diagram with 3D direct numerical simulation results, and make a comparison with the linear stability analysis.

The theoretical division of (in)stability regimes in a Hele-Shaw cell is based on the local capillary number at the inlet, $Ca_{in}$. The critical capillary number $Ca_c$ for a particular angled Hele-Shaw cell is obtained from Eq.~\eqref{eq:Cac}. By setting $Ca_{in}=Ca_c$ or $Ca_{out}=Ca_c$, we obtain the $Ca_{in}$ and $Ca_{out}$ values at the onset of the stability transition, respectively. Then, we refer both critical cases back to $Ca_{in}$, by computing the corresponding value of $Ca_{in}$ for $Ca_{out}=Ca_c$. Thus, we obtain two sets of $Ca_{in}$ values, computed from setting $Ca_{in}=Ca_c$ and from $Ca_{out}=Ca_c$, respectively, which divide the flow into three regimes in the $(\alpha,Ca_{in})$ plane. Note that the stability diagram depends on the Hele-Shaw cell's  geometry because the $Ca_{in}$ that we compute from $Ca_{out}=Ca_c$ is related to the length of the channel. However, in our simulation the interface never reaches the end of the Hele-Shaw cell. Therefore, the horizontal length scale to be used in the nondimensionalization should be reconsidered to determine whether the interface will change its stability during its transit of the initial length of the Hele-Shaw cell. To make a reasonable comparison with the simulation results, we evaluate $Ca_{out}$ at $x = 40$ mm, the maximum distance reached by the interfaces in the simulations.

The theoretically predicted regime divisions are shown in Fig.~\ref{fig:StabilityDiagram} as dashed curves. The $Ca_{in}$ values calculated from $Ca_{in} = Ca_c$ are plotted with circles, while the ones obtained from $Ca_{local} = Ca_c$ are plotted with crosses. 
Note that there is an intersection between the two curves at $\alpha =0$ because in this case there is no sense in which to distinguish $Ca_{in}$ from $Ca_{local}$.  For $\alpha\ne0$, the local capillary number at the inlet $Ca_{in}$ is the minimum capillary number in converging cells ($\alpha < 0$), while it is the maximum capillary number in diverging cells ($\alpha > 0$). Therefore, the critical $Ca_{in}$ from $Ca_{in} = Ca_c$ separates Regime II from Regime III in converging cells, while it separates Regime I from Regime II in diverging cells. The intersection could also be interpreted as follows: as $|\alpha|$ decreases, the range of Regime II narrows, finally collapsing to a point for $\alpha = 0$ (parallel cell, i.e., the ``classical'' Saffman--Taylor case).

Solid curves are the results of direct numerical simulations, representing the division of the parameter space into three regimes as predicted by the 3D evolution of the interface. Numerical experiments were carried out for $\alpha = 0,~\pm 5 \times 10^{-4}$ and $\pm 10 \times 10^{-4}$, across multiple $Ca_{in}$ values ranging from $0.0067$ to $0.02$ with a step of $0.0013$. The numerical results show a general agreement with the theoretically predicted regime map, except for Regime II. This is not surprising because the marginally stable regime is hard to capture in simulations.

The stability diagram in Fig.~\ref{fig:StabilityDiagram} also captures the effect of depth gradient on the regimes: as $\alpha$ increase, the regimes boundaries move up to larger $Ca_{in}$ values. In particular, we observe that simulations predict a much weaker dependence on $\alpha$ than the linear stability analysis.

\section{Conclusions}
\label{sec:conclusion}

In this work, we analyzed the instability of the interface between immiscible viscous fluids in angled Hele-Shaw cells with small constant depth gradient in the flow-wise direction. We performed a linear stability analysis, and we derived an analytical solution for the time-dependent (not constant) growth rate $\dot{\lambda}(t)$ of perturbations to a flat interface. Our theoretical result takes into account the geometric variations and the local capillary number $Ca$ due to the changing cross-sectional area of the flow conduit. Specifically, we have analyzed the case, not previously considered, of the interface instability developing on a time scale on which the flow-wise geometric variations matter. Consequently, dynamic changes in stability were shown to be possible during the interface's propagation. Then, we obtained a critical capillary number $Ca_c$ by letting $\dot{\lambda}=0$. By comparing the local capillary number with $Ca_c$, we put forward a division of the stability diagram into three regimes. In Regime I, the interface is always stable; in Regime II, the interface remains neutrally stable (in a parallel cell), while in angled cells, an exchange of stability occurs at a specific position in the cell: from stable to unstable (in a converging cell), or from unstable to stable (in a diverging cell). This regime classification implies, in particular, that whether or not a depth gradients is stabilizing (or destabilizing) for a given inlet flow rate (thus, given capillary number) depends on which regime the flow falls into. Therefore, a negative depth gradient (converging geometry) is not generically a stabilizing mechanism.

Next, we performed 3D direct numerical simulations of the same physical system, using the interFoam solver built onto the OpenFOAM{\textsuperscript\textregistered} platform, in order to verify the proposed three regimes theory. In Regime III, the finger's length and growth rate computed from simulations agree well with theoretical predictions, verifying the linear stability analysis for the classical Saffman--Taylor instability. Meanwhile, in Regimes I and II, the simulation results support the regimes' existence, but do not match the $\dot{\lambda}$ values predicted by linear stability, due to the former's limitations discussed above.

More importantly, however, the proposed linear theory, when compared to the simulation results, correctly captures the general dependence of the interface growth rate on the depth gradient. Specifically, in Regime I, simulation results show that the integral of the growth rate $\lambda$ decreases slightly with the gap gradient; meanwhile, linear stability analysis exaggerates the effect, which is expected to be corrected by employing a weakly-nonlinear stability analysis. Simulations also suggest that the suppression of fingering in Regime I is most robust in diverging angled Hele-Shaw cells. Furthermore, in Regime II, the effect of the gradient is stronger than that in Regime I. The gap gradient's effect is significant in Regime III, in which case it acts to trigger the fingering instability, especially in converging cells. In all three regimes, the diverging (converging) cells have a relatively stabilizing (destabilizing) effect, with respect to the interface evolution in parallel cells for the same $Ca$ value, which is contrary to the intuition developed in recent experimental studies. 

Finally, we compared the regime stability diagram in the $(\alpha, Ca_{in})$ plane as predicted by theory and as computed from simulations. Although there are some expected systematic sources of error between theory and simulation here, the stability diagram in Fig.~\ref{fig:StabilityDiagram}, to the best of our knowledge, for the first time, compares 3D direct numerical simulations with linear stability analysis of instability in an angled Hele-Shaw cell. Consequently, the present work might provide a framework through which to understand interfacial (in)stability in the presence of geometry variations. Specifically, researchers may:
\begin{enumerate}
    \item compare quantitatively their simulation or experimental results against the stability diagram (Fig.~\ref{fig:StabilityDiagram}) to determine in which regime applies, and therefore understand whether the depth gradient is stabilizing or destabilizing;
    \item solve the ordinary differential equation~\eqref{eq:GR} for the growth of the interface under the specific flow and geometric conditions of their simulation or experiment, to determine the expected behavior of perturbations of the fluid--fluid interface (growth, decay, or \emph{both}).
\end{enumerate}

Additionally, we hope that this stability diagram provides a methodology for addressing the question of how one might improve the sweep efficiency of fluid--fluid displacements in complex fractured rock composed of a network of non-uniform passages. For example, if the flow (for a particular diverging or converging flow passage) can be controlled and forced into Regime I, or at least into Regime II, the interfacial instabilities can be mitigated. Such fundamental understanding of fingering control can also be employed to minimize the risk of geomechanical phenomena during overflushing. In fact, it was shown \citep{OBZD18} that viscous fingering (occurring in Regime III) causes a non-uniform sweep of proppants in a fracture, which are important to distribute uniformly to prevents the fracture's collapse.

Future work should consider a weakly-nonlinear analysis with mode coupling to extend the predictive capability of the stability theory. Furthermore, direct numerical simulation could be employed to verify studies on the critical wave number $k_c$ at fixed $Ca$, such as the prior work of \citet{Miranda1998}. In fact, simulations possess a crucial advantage over experiments: simulations allow precise control over the initial conditions, including the wave number of the interfacial disturbance. Finally, the effects of shear-dependent viscosity \citep{WXLPZ19}, yield stress \citep{BOD15} and viscoelasticity \citep{MS14} on the interfacial instability in angled Hele-Shaw cells should be investigated, building upon the previous work in parallel cells. This aspect of future work is particularly relevant given that non-Newtonian fluids are commonly used in hydraulic fracturing applications \citep{Barbati2016,O17}.

\subsection*{Acknowledgements}

Acknowledgment is made to the donors of the American Chemical Society Petroleum Research Fund for support of this research under ACS PRF award \# 57371-DNI9. I.C.C.\ would like to thank T.~T.~Al-Housseiny for many enlightening discussions on control of interfacial instabilities over the years. We thank Michael Jackson (Queensland University of Technology) for pointing out a missing term in an earlier version of the expression for the growth rate. 


\appendix
\section*{Appendix}

To derive the pressure jump at the interface, we first substitute Eq.~\eqref{eq:pj} into Eq.~\eqref{eq:govern3} and collect terms at $\mathcal{O}(1)$ to obtain
\begin{equation}
\frac{d^2p_{0j}}{dx^2}+\frac{3\alpha}{h_{in}}\frac{dp_{0j}}{dx}=0, \quad \quad j=1,2\qquad (\alpha\ll1).
\label{eq:p0}
\end{equation}
Solving the ordinary differential equation in Eq.~\eqref{eq:p0}, we have
\begin{equation}
p_{0j} = C_{1j} e^{-{3 \alpha}x/h_{in}} + C_{2j}. \label{eq:p0sol}
\end{equation}
Here, the constants $C_{2j}$ are set by the arbitrary pressure gauge for each fluid; specifically, we can set $C_{2j}=0$ without loss of generality. (Eq.~\eqref{eq:govern} can also be solved without the linearization in $\alpha$ that leads to Eq.~\eqref{eq:p0} \citep{P60}. However, it is not clear whether the significantly more complicated expressions contribute much within the lubrication approximation with $\alpha\ll1$.) Now, we must specify boundary conditions.  At the interface, $x=\zeta =\zeta_0(t) + \mathcal{O}(\epsilon)$, the $x$-velocities at the leading order (i.e., for an unperturbed interface) must match:
\begin{equation}
\lim_{x\to\zeta_0^-} u_{x1} = \lim_{x\to\zeta_0^+} u_{x2}= U\big(\zeta_0(t)\big), \label{eq:p0bc}
\end{equation}
where $U$ is the local velocity at the interface. Then, using Eqs.~\eqref{eq:Darcy} and \eqref{eq:p0sol}, we can re-express Eq.~\eqref{eq:p0bc} as
\begin{subequations}
\begin{align}
\lim_{x\to\zeta_0^-} \frac{dp_{01}}{dx} &= -\left.\frac{12\mu_1 U(x)}{[h(x)]^2}\right|_{x=\zeta_0(t)},\\
\lim_{x\to\zeta_0^+} \frac{dp_{02}}{dx} &= -\left.\frac{12\mu_2 U(x)}{[h(x)]^2}\right|_{x=\zeta_0(t)}.
\end{align}
\label{eq:p0_BC}
\end{subequations}
Applying the boundary condition from Eqs.~\eqref{eq:p0_BC} to the solution in Eq.~\eqref{eq:p0sol}, we have
\begin{equation}
p_{0j} = \frac{4\mu_j U\big(\zeta_0(t)\big) h_{in}}{ \alpha [h\big(\zeta_0(t)\big)]^2} e^{-\frac{3\alpha}{h_{in}}\big(x-\zeta_0(t)\big)} \qquad (\alpha\ll1). \label{eq:p0sol2}
\end{equation}

Similarly, substituting Eqs.~\eqref{eq:pj} and \eqref{eq:p1_s} into Eq.~\eqref{eq:govern3} and collecting terms at $\mathcal{O}(\epsilon)$, we have 
\begin{equation}
\frac{d^2g_{j}}{dx^2}+\frac{3\alpha}{h_{in}}\frac{dg_{j}}{dx}-k^2g_{j}=0,\qquad j=1,2 \qquad (\alpha\ll1),
\label{eq:gj}
\end{equation}
subject to
\begin{equation}
\lim_{x \to -\infty} g_{1}(x) = \lim_{x \to +\infty} g_{2}(x) = 0. \label{eq:g_bc}
\end{equation}
The solution to Eq.~\eqref{eq:gj} is of the form $g_{j} = b_{j1}e^{m_1x} + b_{j2}e^{m_2x}$, where $b_{j1}$ and $b_{j2}$ are constants, and
\begin{equation}
m_{1,2} = -\frac{3\alpha}{2h_{in}} \left(1 \pm \sqrt[]{1+\frac{4k^2 h_{in}^2}{9\alpha^2}} \,\right). 
\label{eq:m}
\end{equation}
Therefore,
\begin{equation}
g_{1}(x) = b_{11} e^{m_1 x} + b_{12} e^{m_2 x}.
\end{equation}

In a diverging cell, $\alpha > 0$, thus $m_1<0$ and $m_2>0$. From the boundary condition from Eq.~\eqref{eq:g_bc}, $b_{11}=0$. In a converging cell, $\alpha < 0$, thus $m_1>0$, $m_2<0$ and $b_{12} = 0$. Hence,
\begin{equation}
g_{1}(x) = 
\begin{cases}
b_{12} e^{m_2 x}, \quad &\alpha > 0;\\
b_{11} e^{m_1 x},  \quad &\alpha < 0. 
\end{cases}
\label{eq:g_1n}
\end{equation}
Similarly, 
\begin{equation}
g_{2}(x) = 
\begin{cases}
b_{21} e^{m_1 x}, \quad &\alpha > 0;\\
b_{22} e^{m_2 x},  \quad &\alpha < 0. 
\end{cases}
\label{eq:g_2n}
\end{equation}
Following \citep{Al-Housseiny2013}, Eq.~\eqref{eq:g_1n} and \eqref{eq:g_2n} can be combined into a single equation:
\begin{equation}
g_{j} = \hat{b}(j,\alpha) e^{\hat{m}(j,\alpha) x},
\end{equation}
where 
\begin{equation}
\hat{m}(j,\alpha) 
= -\frac{3\alpha}{2h_{in}} \left(1 + (-1)^j \sgn(\alpha) \, \sqrt[]{1+\frac{4k^2 h_{in}^2}{9\alpha^2}} \,\right),
\label{eq:mhat_j_alpha}
\end{equation}
where $\sgn(\alpha) = |\alpha|/\alpha$ for $\alpha \ne 0$ and vanishes otherwise. Therefore, Eq.~\eqref{eq:p1_s} becomes
\begin{equation}
p_{1j} = \hat{b}(j,\alpha) e^{\hat{m}(j,\alpha)x + iky + \lambda(t)} \qquad (\alpha\ll1).
\label{eq:p1sol}
\end{equation}
Finally, substituting the leading-order pressure from Eq.~\eqref{eq:p0sol2} and the pressure-correction term from Eq.~\eqref{eq:p1sol} into the definition of $p_j$ from Eq.~\eqref{eq:pj}, leads to 
\begin{equation} 
p_j(x,y,t)= \frac{4\mu_j U\big(\zeta_0(t)\big) h_{in}}{ \alpha [h\big(\zeta_0(t)\big)]^2} e^{-\frac{3\alpha}{h_{in}}\big(x-\zeta_0(t)\big)} + \epsilon \hat{b}(j,\alpha) e^{\hat{m}(j,\alpha)x + iky + \lambda (t)}. \label{eq:pj2}
\end{equation}

Now, onto the boundary conditions for $p_{j}$ at the interface. First, consider the kinematic condition, which states that interface velocity equals the fluid velocity at the interface:
\begin{equation}
\frac{\partial \zeta}{\partial t} = \bm{u}_j \cdot \hat{\bm{n}} |_{x=\zeta(y,t)},
\label{eq:kine0}
\end{equation}
where $\bm{u}_j$ is given by Eq.~\eqref{eq:Darcy}. Letting $F=0$, where $F(x,y) := x - \zeta(y,t)$, implicitly define the interface position, the unit surface normal $\hat{\bm{n}}$ can be defined, in Cartesian coordinates, as
\begin{equation}
\hat{\bm{n}} = \frac{\nabla F}{\|\nabla F\|} = \frac{1}{\|\nabla F\|}\left( \frac{\partial F}{\partial x}, \frac{\partial F}{\partial y} \right)
= \left[ 1 
+ \left( \frac{\partial \zeta}{\partial y} \right)^{2} \right]^{-1/2} \left( 1, -\frac{\partial \zeta}{\partial y} \right).
\end{equation}
Using a Taylor-series expansion for $|\partial\zeta/\partial y| \ll 1$, we have 
\begin{equation}
\begin{split}
\hat{\bm{n}} &= \left( 1, -\frac{\partial \zeta}{\partial y} \right) 
\left[  1 - \frac{1}{2} \left( \frac{\partial \zeta}{\partial y} \right)^2 + \mathcal{O} \left( \left( \frac{\partial \zeta}{\partial y} \right)^4 \right) \right] \\
&\simeq \left( 1,  -\epsilon a e^{iky+\lambda (t)} ik \right),
\label{eq:n_zeta_expand}
\end{split}
\end{equation}
if only leading-order terms are kept.

Next, we combine Eq.~\eqref{eq:Darcy} with $p_j$ as given by Eq.~\eqref{eq:pj2} to obtain the Darcy velocities
\begin{equation}
\bm{u}_j = -\frac{[h\big(\zeta(t)\big)]^2}{12 \mu_j} \left( \frac{\partial p_j}{\partial x}, \frac{\partial p_j}{\partial y} \right), \quad \quad j=1,2.
\label{eq:u_j_expand_iface}
\end{equation}
Combining Eqs.~\eqref{eq:n_zeta_expand} and \eqref{eq:u_j_expand_iface}, we find the normal velocity to be
\begin{equation}
\bm{u}_j \cdot \hat{\bm{n}} 
 = -\frac{[h\big(\zeta(t)\big)]^2}{12 \mu_j} \left(\frac{\partial p_j}{\partial x} - \frac{\partial p_j}{\partial y} \frac{\partial \zeta}{\partial y}\right). 
\label{eq:u*n}
\end{equation}
Therefore, each fluid's velocity normal to the interface is given by 
\begin{equation}
\begin{split}
\bm{u}_j \cdot \hat{\bm{n}} |_{x=\zeta(y,t)} & = U\big(\zeta_0(t)\big) + \epsilon \left\{ U\big(\zeta_0(t)\big)a \left( \frac{2\alpha  }{h\big(\zeta_0(t)\big) } - \frac{3\alpha }{h_{in}} \right) -\frac{[h\big(\zeta_0(t)\big)]^2 \hat{b} \hat{m}}{12 \mu_j} e^{\hat{m} \zeta_0}\right\} e^{iky + \lambda (t)} + \mathcal{O}(\epsilon^2). \label{eq:u*n2}
\end{split}
\end{equation}

Substituting Eq.~\eqref{eq:iface-perturb2} and Eq.~\eqref{eq:u*n2} back into Eq.~\eqref{eq:kine0}, we obtain
\begin{equation}
\begin{split}
\hat{b}(j,\alpha) &=  -\frac{12 \mu_j U\big(\zeta_0(t)\big) a}{\hat{m} [h\big(\zeta_0(t)\big)]^2} \left[ \frac{\dot{\lambda}}{U\big(\zeta_0(t)\big)}  + \alpha \left(\frac{3}{h_{in}} - \frac{2}{h\big(\zeta_0(t)\big)} \right)  \right] e^{-\hat{m}\zeta_0(t)}.
\end{split}
\end{equation}
For sufficiently large wave numbers compared to the gap gradient \citep{Al-Housseiny2013}, i.e., $|kh_{in}/\alpha| \gg 1$, we can approximate the exponents, given in Eq.~\eqref{eq:mhat_j_alpha}, as $\hat{m}(1,\alpha) \approx k$ and $\hat{m}(2,\alpha) \approx -k$. Then, the pressure at the interface becomes
\begin{multline}
p_j(x,y,t)|_{x=\zeta} = \frac{4\mu_j U\big(\zeta_0(t)\big) h_{in}}{ \alpha [h(\zeta_0(t))]^2}\\ - \epsilon e^{iky+\lambda (t)} \frac{12 U\big(\zeta_0(t)\big) a}{[h\big(\zeta_0(t)\big)]^2}\left\{\mu_j +\left[ \frac{\dot{\lambda}}{U\big(\zeta_0(t)\big)}  + \alpha \left(\frac{3}{h_{in}} - \frac{2}{h\big(\zeta_0(t)\big)} \right)  \right] \frac{\mu_j}{\hat{m}(j,\alpha)} \right\} + \mathcal{O}(\epsilon^2).
\end{multline}
Thus, finally, the pressure difference across the fluid--fluid interface takes the form of Eq.~\eqref{eq:p2-p1} above.

\end{document}